\documentclass[10pt,conference]{IEEEtran}
\IEEEoverridecommandlockouts
\usepackage{cite}
\usepackage[T1]{fontenc}
\usepackage{amsmath,amsthm,amssymb,amsfonts,braket}
\usepackage{todonotes}
\usepackage{algorithmicx}
\usepackage{graphicx}
\usepackage{textcomp}
\usepackage{xcolor}
\usepackage{algorithm,balance}
\usepackage[noend]{algpseudocode}
\usepackage{breqn}
\usepackage{physics}
\usepackage{subfigure}
\usepackage{xcolor}
\usepackage{mathtools}
\usepackage{lipsum}

\usepackage{tikz}
\usetikzlibrary{decorations.pathreplacing,angles,quotes,calc, decorations.pathmorphing, decorations.shapes}
\usepackage{soul}
\usepackage{subcaption}
\usepackage{phoenician}

\newtheorem {theorem} {Theorem}

\newcommand{\kb}[1]{\ket{#1}\bra{#1}}

\newcommand\note[3]{{\textcolor{#1}{[\textsf{#2}: #3]}}}

\newcommand{\bing}[1]{\note{blue}{BW}{#1}}

\newcommand{\sam}[1]{\note{blue}{SO}{#1}}

\algnewcommand\algorithmicforeach{\textbf{for each}}
\algdef{S}[FOR]{ForEach}[1]{\algorithmicforeach\ #1\ \algorithmicdo}

\def\BibTeX{{\rm B\kern-.05em{\sc i\kern-.025em b}\kern-.08em
    T\kern-.1667em\lower.7ex\hbox{E}\kern-.125emX}}

\graphicspath{ {./images/} }

\makeatletter
\newcommand{\linebreakand}{%
  \end{@IEEEauthorhalign}
  \hfill\mbox{}\par
  \mbox{}\hfill\begin{@IEEEauthorhalign}
}
\makeatother

\usepackage{epsfig}
\usepackage{algorithm}

\newcommand{\remove}[1]{}

\newcommand{\mysection}[1]{\vspace{-.05in}\section{#1}\vspace{-.02in}}
\newcommand{\mysubsection}[1]{\vspace{-.05in}\subsection{#1}\vspace{-.02in}}

\newtheorem{lemma}{Lemma}
\newtheorem{corollary}{Corollary}

\newcommand{\ls}[1]
   {\dimen0=\fontdimen6\the\font
    \lineskip=#1\dimen0
    \advance\lineskip.5\fontdimen5\the\font
    \advance\lineskip-\dimen0
    \lineskiplimit=.9\lineskip
    \baselineskip=\lineskip
    \advance\baselineskip\dimen0
    \normallineskip\lineskip
    \normallineskiplimit\lineskiplimit
    \normalbaselineskip\baselineskip
    \ignorespaces
   }

\newcommand{\singlefig}[3]{
\begin{figure}
\centerline{
    \setlength{\epsfysize}{0.23\textwidth}
 \epsffile{\Figdir#1}
} \vspace{-0.05in} \caption{\small #2} \label{fig:#3}
\end{figure}
}

\newcommand{\tydubfigsingle}[6]{
\begin{figure}
\centerline{
    \begin{minipage}{0.30\textwidth}
      \begin{center}
        \leavevmode
        \setlength{\epsfxsize}{0.85\textwidth}
        \setlength{\epsfysize}{0.72\textwidth}
        \epsffile{\Figdir#1}
       \newline{\small (a) #2}
      \end{center}
    \end{minipage}
    \hspace*{-1cm}
    \begin{minipage}{0.30\textwidth}
      \begin{center}
        \leavevmode
        \setlength{\epsfxsize}{0.85\textwidth}
        \setlength{\epsfysize}{0.72\textwidth}
        \epsffile{\Figdir#3}
       \newline{\small (b) #4}
      \end{center}
    \end{minipage}
} \caption{\small #5}\label{fig:#6}
\end{figure}
}

\newcommand{\triplefignostretch}[8]{
\begin{figure*}[t]
\centerline{
    \begin{minipage}{0.30\textwidth}
      \begin{center}
        \leavevmode
        \setlength{\epsfxsize}{\textwidth}
        \epsffile{\Figdir#1}\\
       {\small (a) #2}
      \end{center}
    \end{minipage}
    \hspace{0.2in}
    \begin{minipage}{0.30\textwidth}
      \begin{center}
        \leavevmode
        \setlength{\epsfxsize}{\textwidth}
        \epsffile{\Figdir#3}\\
       {\small (b) #4}
      \end{center}
    \end{minipage}
    \hspace{0.2in}
    \begin{minipage}{0.30\textwidth}
      \begin{center}
        \leavevmode
        \setlength{\epsfxsize}{\textwidth}
         \epsffile{\Figdir#5}\\
       {\small (c) #6}
      \end{center}
    \end{minipage}
} \caption{\small #7}\label{fig:#8}
\end{figure*}
}

\newcommand{\triplefigmed}[8]{
\begin{figure*}[t]
\centerline{
    \begin{minipage}{0.30\textwidth}
      \begin{center}
        \leavevmode
        \setlength{\epsfxsize}{0.75\textwidth}
        \epsffile{\Figdir#1}\\
       {\small (a) #2}
      \end{center}
    \end{minipage}
    \hspace{0.2in}
    \begin{minipage}{0.30\textwidth}
      \begin{center}
        \leavevmode
        \setlength{\epsfxsize}{0.75\textwidth}
        \epsffile{\Figdir#3}\\
       {\small (b) #4}
      \end{center}
    \end{minipage}
    \hspace{0.2in}
    \begin{minipage}{0.30\textwidth}
      \begin{center}
        \leavevmode
        \setlength{\epsfxsize}{0.75\textwidth}
         \epsffile{\Figdir#5}\\
       {\small (c) #6}
      \end{center}
    \end{minipage}
} 
\vspace{-0.1in}\caption{\small #7}\label{fig:#8}
\end{figure*}
}

\newcommand{\Figdir}{./}

\begin{document}


\title{Efficient Multiparty Quantum Key Distribution over Quantum Networks}


\author{
\IEEEauthorblockN{Samuel Oslovich\IEEEauthorrefmark{1},  Bing Wang\IEEEauthorrefmark{1}, Walter O. Krawec\IEEEauthorrefmark{1}, Kenneth Goodenough\IEEEauthorrefmark{2}}

\IEEEauthorblockA{\IEEEauthorrefmark{1}School of Computing, University of Connecticut, Storrs, CT, USA} 
\IEEEauthorblockA{\IEEEauthorrefmark{2}Manning College of Information \& Computer Sciences, University of Massachusetts, Amherst, MA, USA}
}

\maketitle

\begin{abstract}
Multiparty quantum key distribution (QKD) is useful for many applications that involve secure communication or collaboration among multiple parties.
While it can be achieved using pairwise QKD, a more efficient approach is to achieve it using multipartite entanglement distributed over quantum networks that connect the multiple parties. Existing studies on multipartite entanglement distribution, however, are not designed for multiparty QKD, and hence do not aim to maximize secret key generation rate. In this paper, we design efficient strategies for multiparty QKD over quantum networks. For 3-party QKD,  we derive closed-form expressions for analyzing key distribution over quantum networks.  We then use it to develop an efficient strategy for 3-party QKD by packing multiple stars 
that connect the 3 parties. 
For the general form of $N$-party QKD, we develop an approach that packs multiple trees 
to connect the $N$ parties, while 
directly incorporating the estimated key rates on network paths. Extensive evaluation of our strategies, in both grid and random graphs, under a wide range of settings, demonstrates that our schemes achieve high key rate, which degrades gracefully when increasing the number of parties.   
\end{abstract}

\section{Introduction} \label{sec:intro}

Quantum key distribution (QKD) leverages the principles of quantum mechanics to allow two parties, Alice and
Bob, to establish a secret key that
is information theoretically secure~\cite{Pirandola19:QKD-Survey}. 
Multiparty QKD, also called {\em conference key agreement}, extends  pairwise QKD to allow multiple parties to establish a common  information-theoretic secret key.
It  is  useful for many applications, including 
secure multiparty communication, distributed cryptographic applications, and quantum secure multi-party computation~\cite{Murta20:quantum-CKA}.
When the parties are at geographically distributed locations, long-distance quantum networks, as envisioned in the ``quantum internet''~\cite{kimble2008quantum,caleffi2018quantum,wehner2018quantum}, provide a mechanism to connect the multiple parties using intermediate quantum repeaters.
Specifically, the usage of quantum repeaters addresses the issue of the exponential decay in the rate as a function of distance,
allowing higher entanglement generation rates, and hence higher key generation rate. 
While multiparty QKD can be achieved through pairwise QKD, which first establishes pairwise secret keys and then uses secure communication to establish a common key among all the parties, this process is inefficient~\cite{Epping_2017}. A more efficient alternative is distributing multipartite entanglement directly to the multiple parties over a quantum network, which allows the parties to leverage the multipartite entangled state, to achieve a higher key rate. 


Existing experimental studies~\cite{Proietti21:CKA,Pickston_2023} have demonstrated the feasibility of multiparty QKD using multipartite entanglement distribution over small-scale quantum networks.
Efficient multiparty QKD over large-scale quantum networks, however, faces many challenges. These challenges are due to limitations of near-term technologies, such as lossy and noisy quantum channels, limited quantum memory with short lifetime, and probabilistic  entanglement swapping operations. To improve the key rate of multiparty QKD, an important building block is efficient multipartite entanglement distribution over quantum networks. While this problem has been explored in recent studies~\cite{Meignant19:graph-state,Fischer21:graph-state,Bugalho_2023,Sutcliffe_2023}, their designs do not target multiparty QKD.
Specifically, the studies in~\cite{Meignant19:graph-state,Fischer21:graph-state} aim to minimize the number of consumed entangled pairs  under idealized conditions, with no consideration of loss and noise. The study in~\cite{Bugalho_2023} considers a more realistic noisy 
quantum network setting, and develops a multi-objective
routing algorithm considering both entanglement generation rate and  fidelity. 
The study in~\cite{Sutcliffe_2023} develops multipath routing techniques to improve entanglement throughput, with no consideration of  quantum link fidelity, which is important for quantum network applications, especially QKD where key rate is affected  heavily by fidelity.

In this paper, we develop efficient multipartite entanglement distribution strategies with the aim of maximizing key rate for 
multiparty QKD.
Our approach incorporates the various constraints in near-term quantum networks and is applicable to general quantum network topologies.  We start with the simplest form of multiparty QKD, 3-party QKD. In this setting, we derive closed-form expressions for estimating errors required for obtaining key rate.
We then use them to develop an efficient strategy for 3-party QKD by packing multiple star topologies that connect the three parties. 
For the general form of $N$-party QKD, we develop an approach that packs multiple tree topologies to connect the $N$ parties, while 
directly incorporating the estimated key rates on network paths into the design. While our approach also leverages multipath routing as in~\cite{Sutcliffe_2023}, the multipartite entanglement distribution strategy that we design differs significantly from that in~\cite{Sutcliffe_2023} in that our design explicitly considers fidelity and has the goal of maximizing key rate for multiparty QKD. 

We evaluate our proposed approach in both grid and random graphs under a wide range of settings. The evaluation results show that our proposed approach achieves high key rate, and degrades gracefully when increasing the number of parties. Furthermore, it significantly outperforms several baselines. Specifically, in the settings we explore, we find that our dynamic multi-tree algorithm achieves up to $223\%$ higher key rate compared to a fixed multi-tree algorithm. 


The rest of the paper is organized as follows. In Section \ref{sec:background}, we present background material on multiparty QKD and network model. In Sections \ref{sec:3-party} and \ref{sec:n-party}, we present our designs for 3-party and $N$-party QKD, respectively. In Section \ref{sec:eval}, we present our evaluation results.  In Section \ref{sec:related}, we briefly review related work. Last, Section~\ref{sec:concl} concludes the paper.
\section{Background and Network Model} \label{sec:background}
In this section, we briefly overview the multipartite QKD protocol that is used in this study, and present the network model. 

\begin{figure}[t]
    \def\qubitSize{0.1}
    \definecolor{electricpurple}{rgb}{0.75, 0.0, 1.0}
    \centering
    \begin{minipage}{0.49\textwidth}
    \centering
    
    \def\qubitSize{0.1}
\definecolor{electricpurple}{rgb}{0.75, 0.0, 1.0}
\begin{tikzpicture}[scale=0.92]
    \draw[draw=black] (-0.8,1.3) rectangle (3,3);
    \filldraw[fill=blue] (-0.4, 2.7) circle (\qubitSize);
    \node[] at (1.4-0.75, 2.7) {\textbf{Qubit}};
    \draw[decorate, decoration={coil,aspect=0}, line width=0.5mm, draw=electricpurple,segment length=12pt] (0.1-0.8, 2.2) -- (0.8- 0.8, 2.2);
    \node[] at (2.2-0.8, 2.15) {\textbf{Entanglement}};
    \draw[dashed, line width=0.5mm] (0.4-0.8, 1.6) circle (0.2);
    \node[] at (2.1-0.7, 1.6) {\textbf{Measurement}};

    \node[shape=circle,draw=black,minimum size=1cm, line width=0.25mm] (N1) at (0,0) {};
    \node[shape=circle,draw=black,minimum size=1cm, line width=0.25mm] (N2) at (3,0) {};
    \node[shape=circle,draw=black,minimum size=1cm, line width=0.25mm] (N3) at (6,0) {};
    \node[shape=circle,draw=black,minimum size=1cm, line width=0.25mm] (N4) at (4.5,1.65) {};
    \node[shape=circle,draw=black,minimum size=1cm, line width=0.25mm] (N5) at (4.5,-1.65) {};

    \node[] at ($(N1)+(0,0.7)$) {\textbf{A}};
    \node[] at ($(N2)+(0,0.7)$) {\textbf{B}};
    \node[] at ($(N3)+(0,0.7)$) {\textbf{C}};
    \node[] at ($(N4)+(0.7,0)$) {\textbf{D}};
    \node[] at ($(N5)+(0.7,0)$) {\textbf{E}};

    \filldraw[fill=blue] ($(N1)+(0.2,0)$) circle (\qubitSize);

    \filldraw[fill=blue] ($(N2)-(0.2,0)$) circle (\qubitSize);
    \filldraw[fill=blue] ($(N2)+(0.2,0)$) circle (\qubitSize);

    \filldraw[fill=blue] ($(N3)-(0.2,0)$) circle (\qubitSize);
    \filldraw[fill=blue] ($(N3)+(0,0.25)$) circle (\qubitSize);
    \filldraw[fill=blue] ($(N3)-(0,0.25)$) circle (\qubitSize);
    
    \filldraw[fill=blue] ($(N4)-(0,0.2)$) circle (\qubitSize);

    \filldraw[fill=blue] ($(N5)+(0,0.2)$) circle (\qubitSize);

    \draw[decorate, decoration={coil,aspect=0}, line width=0.5mm, draw=electricpurple,segment length=12pt] ($(N1)+(0.2,0)$) -- ($(N2)-(0.2,0)$);
    \draw[decorate, decoration={coil,aspect=0}, line width=0.5mm, draw=electricpurple,segment length=12pt] ($(N2)+(0.2,0)$) -- ($(N3)-(0.2,0)$);

    \draw[decorate, decoration={coil,aspect=0}, line width=0.5mm, draw=electricpurple,segment length=12pt] ($(N3)+(0,0.25)$) -- ($(N4)-(0,0.2)$);
    \draw[decorate, decoration={coil,aspect=0}, line width=0.5mm, draw=electricpurple,segment length=12pt] ($(N3)-(0,0.25)$) -- ($(N5)+(0,0.2)$);

    \draw[dashed, line width=0.5mm] (N2) circle (0.9);
    \draw[dashed, line width=0.5mm] (N3) circle (0.9);


\end{tikzpicture}

    {\small (a) BSM at node $B$. Entanglement fusion at node $C$.}
    \end{minipage}

    \begin{minipage}{0.49\textwidth}
    \centering
    \vspace{0.3cm}
    \begin{tikzpicture}[scale=0.92]

    \node[shape=circle,draw=black,minimum size=1cm, line width=0.25mm] (N1) at (0,0) {};
    \node[shape=circle,draw=black,minimum size=1cm, line width=0.25mm] (N2) at (3,0) {};
    \node[shape=circle,draw=black,minimum size=1cm, line width=0.25mm] (N3) at (6,0) {};
    \node[shape=circle,draw=black,minimum size=1cm, line width=0.25mm] (N4) at (4.5,1.65) {};
    \node[shape=circle,draw=black,minimum size=1cm, line width=0.25mm] (N5) at (4.5,-1.65) {};

    \node[] at ($(N1)+(0,0.7)$) {\textbf{A}};
    \node[] at ($(N2)+(0,0.7)$) {\textbf{B}};
    \node[] at ($(N3)+(0,0.7)$) {\textbf{C}};
    \node[] at ($(N4)+(0.7,0)$) {\textbf{D}};
    \node[] at ($(N5)+(0.7,0)$) {\textbf{E}};

    \filldraw[fill=blue] ($(N1)+(0.2,0)$) circle (\qubitSize);


    \filldraw[fill=blue] ($(N3)-(0.2,0)$) circle (\qubitSize);
    
    \filldraw[fill=blue] ($(N4)-(0,0.2)$) circle (\qubitSize);

    \filldraw[fill=blue] ($(N5)+(0,0.2)$) circle (\qubitSize);

    \draw[decorate, decoration={coil,aspect=0}, line width=0.5mm, draw=electricpurple,segment length=12pt] ($(N1)+(0.2,0)$) to[in=-140, out=-40] ($(N3)-(0.2,0)$);

    \draw[decorate, decoration={coil,aspect=0}, line width=0.5mm, draw=electricpurple,segment length=12pt] ($(N3)-(0.2,0)$) -- ($(N4)-(0,0.2)$);
    \draw[decorate, decoration={coil,aspect=0}, line width=0.5mm, draw=electricpurple,segment length=12pt] ($(N3)-(0.2,0)$) -- ($(N5)+(0,0.2)$);



\end{tikzpicture}

    {\small (b) Resulting 4-GHZ state shared by $A$, $C$, $D$, and $E$.}
    \end{minipage}

    \caption{\small Illustration of BSM (entanglement swapping) and {entanglement fusion}.
    (a) Node $B$ performs BSM
    on its two qubits, resulting 
    a shared Bell state between $A$ and $C$. Node $C$ performs {entanglement fusion}, resulting in a shared 4-GHZ state between $A$, $C$, $D$, and $E$. (b) The resulting shared 4-GHZ state.}
    \label{fig:bsm}
\end{figure}

\subsection{$N$-party QKD Protocol}  \label{sec:n-QKD}

We use the $N$-BB84 protocol in \cite{MP_QKD}, which is a $N$-partite variant of the BB84 protocol~\cite{BB14:QKD}. This $N$-partite protocol  considers $N$ parties, denoted as Alice $A$, and a set of Bobs, $B_1, \ldots, B_{N-1}$. The goal of the protocol is to establish a common secret key among all $N$ parties, $A, B_1, \ldots, B_{N-1}$. In the rest of the paper, we refer to $A$ as the {\em leader}, since it leads the error correction among the $N$ parties (see below). 

Just like the standard BB84 protocol, this protocol contains two stages: quantum information processing stage and classical information processing stage. In the quantum information processing stage, the protocol starts with  distributing 
multipartite entangled $N$-GHZ states, $\ket{N\text{-GHZ}} = \frac{1}{\sqrt{2}}(\ket{0}^{\otimes N} + \ket{1}^{\otimes N})$ over the 
quantum channel 
to the $N$ parties.  All parties perform local measurements, either in $Z$ or $X$ basis, based on a preshared key, on their respective quantum systems.  The basis choice is biased so that the $X$ basis is chosen with probability $p < 1/2$.  The basis choice will therefore consume $MH(p)$ 
bits of pre-shared key every time the protocol is run, where $M$ is the number of rounds and $H(p)=-p\log p - (1-p) \log (1-p)$ is the binary entropy function.  This key is then refreshed after each protocol run and, so long as the noise is low enough, there will be more secret key bits produced then consumed \cite{MP_QKD}.  In our evaluations, we consider only the asymptotic key-rate, in which case we may simply take $p$ to be arbitrarily close to zero, and, so $H(p)$ is also arbitrarily close to zero.

In the classical information processing stage, the parties reveal a random sample of the collected data over the authenticated classical channel for parameter estimation. Specifically, this process estimates $Q_{A,B_i}$, i.e., the $Z$ basis quantum bit error rate (QBER) between $A$ and $B_i$, $i=1,\ldots,N-1$, as well as $Q_X$, i.e., the $X$ basis QBER across all $N$ parties. 
%
At this point, the raw keys held by the $N$ parties are partially correlated and partially secret. In order to correct the errors in the raw keys, $A$ performs pairwise error correction with each $B_i$ to create shared raw keys among the $N$ parties. At last, through privacy amplification, the shared raw key is turned into shared secret key for the $N$ parties. The length of the secret key depends on the error rates (obtained using parameter estimation) and the desired level of security. 

Let $r$ be the key rate, i.e., the length of the secret key divided by the number of $N$-GHZ states distributed among the $N$ parties. The study in \cite{MP_QKD} presents a finite-key expression for $r$, which converges to the following asymptotically
\begin{equation}
    \label{eq:n-party-keyrate}
    r = 1- H(Q_X ) - \max_i H(Q_{A,Bi})\ ,
\end{equation}
where $H(\cdot)$ is 
the binary entropy function.  In this paper, we focus on asymptotic results, and use the above asymptotic key rate above.  Later, we will also be interested in the \emph{effective key rate} which is defined to be the number of secret key bits divided by the total timesteps.  This can be derived from the above, by multiplying with a suitable scaling factor based on the number of GHZ states per time step.

\mysubsection{Network Model}\label{sec:model}


The above $N$-party BB84 protocol requires $N$-GHZ states to be distributed to the $N$ parties before running the protocol. We assume the $N$-GHZ states are distributed using a quantum network that connects the $N$ parties.  
In the following, we present the network model.

Consider a quantum network that contains a set of nodes and edges. Among them, the $N$ nodes that will perform $N$-party QKD are referred to as {\em end users} or {\em terminal nodes}, denoted as $T_1, \ldots, T_N$,  and the rest are quantum repeaters. 
%
%
%
As in~\cite{pant2019routing,Shi20:concurrent,amer2020:efficient}, we assume that the nodes have 
synchronized clocks and the network operates in \emph{rounds}. 
Each edge in this network 
represents a fiber link for qubit transmission, which can be noisy and lossy. We assume that all nodes have a quantum memory capable of storing a single qubit for every adjacent node. 
In addition, this qubit can only be stored for a single network round
due to short quantum memory coherence time~\cite{Dahlberg19:link},
after which it must be discarded. The memory may also be noisy, causing the state to decohere with a certain probability.

Each network round is divided into three  \emph{phases}, for completing the quantum information processing stage of $N$-party QKD over the quantum network. 
After a sufficient number of rounds, classical post-processing  is conducted among all the end users to obtain secret key. We next briefly describe these three phases and then  classical post-processing.

\smallskip
\noindent{\em Phase 1.} This phase distributes link-level bipartite entanglement over each link in the network. Specifically, each pair of connected nodes (quantum repeaters), $u$ and $v$, 
attempts to share half of an entangled pair with each other over the fiber link.  This process succeeds with probability $p$, which decays
exponentially with distance~\cite{Svelto10:Principles,Kaushal17:optical,pirandola2017fundamental}. 
In addition, due to noise in the link, memory system, or both, even when entanglement is established successfully, between $u$ and $v$, we assume a depolarizing channel with parameter $\gamma_{u,v}$. That is, the entanglement may depolarize and become a completely mixed state. Thus, at the end of this phase, with probability $p$, each pair of neighbors $u,v$ share the state 
\begin{equation}
  \rho_{u,v} = \gamma_{u,v} \ket{\Phi^+}\bra{\Phi^+} + (1-\gamma_{u,v})\frac{I}{4}\,,
    \label{eq:bell-depolarizing-channel}
\end{equation}
where $\ket{\Phi^+} = \frac{1}{\sqrt{2}}\left(\ket{00}+\ket{11}\right)$ and $I$ is the identity matrix. Otherwise, with probability $1-p$, those neighbors share the vacuum state (i.e., they share no state). We assume nodes are able to determine whether they have a vacuum or not, while they cannot determine if the state is the desired Bell state $\ket{\Phi^+}$ or a completely mixed state. Note that $p$ can vary for different links in a network due to differences in quantum hardware.


\smallskip
\noindent{\em Phase 2}. 
In this stage, a routing algorithm is executed (see Sections~\ref{sec:3-party} and \ref{sec:n-party}) to create $N$-party entanglement among the end users. Specifically, the routing algorithm will determine a routing tree, with end users as leaf nodes or interior nodes of the tree,
and has quantum repeaters as the remaining interior nodes. {The interior nodes in the tree will perform $k$-GHZ projective measurements if $k$ nodes are connected to it. If successful, this will create a long-distance $k$-GHZ state among the $k$ nodes.} 
{In the special case where an interior node is a user, it will perform an entanglement fusion operation to retain entanglement with the resulting state. Entanglement fusion operations are similar to $k$-GHZ projective measurements, but allows for the measuring party to retain a qubit of the entanglement state \cite{de_Bone_2020}.}
Let $q$ denote the probability that the $k$-GHZ state among the $k$ nodes is created successfully, $k\ge 2$.
Note that $q$ can also vary for each node in the network due to differences in quantum hardware.
Henceforth, for simplicity, we consider homogeneous $q$; our approaches in \S\ref{sec:3-party} and \S\ref{sec:n-party} can be used for both homogeneous and heterogeneous $q$.

A branch in the tree is a network path. For a path $P=(u_1,u_2,\ldots,u_{\ell})$ with $\ell$ repeaters, each repeater, $u_i$, will perform a Bell state measurement (BSM) or entanglement swapping 
on its respective half of the shared state. Since each link can only be used once, these network paths cannot have shared links. 
Then with probability $q^\ell$, nodes $u_1$ and $u_\ell$ share a state of the form:
\begin{equation}\label{eq:ms-final-state}
  \rho_{u_1, u_\ell} = \gamma_P \kb{\Phi^+} + \left[1-\gamma_P \right]\frac{I}{4}\,,
\end{equation}
where $\gamma_P=\Pi_{i=1}^{\ell-1} \gamma_{u_i,u_{i+1}}$, and $\gamma_{u_i,u_{i+1}}$ is the depolarizing parameter for link $(u_i,u_{i+1})$. 
An example that illustrates the above BSM and entanglement fusion 
is shown in Fig.~\ref{fig:bsm}.

\smallskip
\noindent{\em Phase 3}. In this phase, with entanglement shared among the $N$ users, each user conducts the quantum portion of the $N$-QKD protocol (see \S\ref{sec:n-QKD}). 
Specifically, each party measures its qubit following the preshared classical key sequence.  At the end of this phase, all remaining qubits in the system are discarded, as we assume quantum memories can only store qubits for a single network round. The network then repeats the above three phases for the next round.

\smallskip
\noindent{\bf Post-processing.} When the network has been running for a sufficient number of rounds, post-processing is executed.  Here, error correction and privacy amplification are run among all end users, generating secret key material.
As we are interested only in the asymptotic performance in this paper,
we assume perfect error correction and use asymptotic key rates for the $N$-party QKD protocol (see \S\ref{sec:n-QKD}).  Our main performance metric is the final {\em secret key generation rate (key rate)} among end users, namely Eq. (\ref{eq:n-party-keyrate}). 

\section{3-Party QKD} \label{sec:3-party}
In this section, we consider the simplest case of $N$-party QKD when $N=3$. We first present 3-GHZ state distribution and then a multipath routing algorithm to maximize the key rate among the 3 parties.  

\subsection{3-GHZ State Distribution}

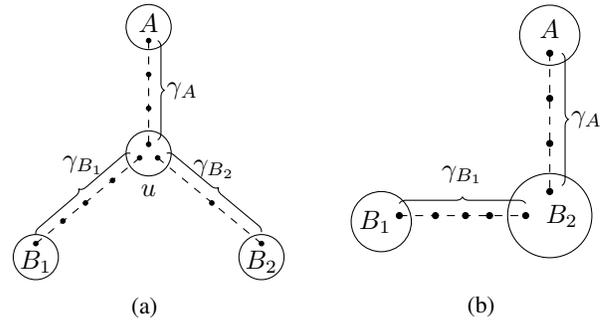
\begin{figure}
    \centering
    \begin{minipage}{0.24\textwidth}
    \centering
        \begin{tikzpicture}[scale=0.6]

            \coordinate (A1) at (4, 6.5);
            \coordinate (B1) at (1.5, 2);
            \coordinate (C1) at (6.5, 2);
            \coordinate (A2) at (4, 4.2);
            \coordinate (B2) at (3.8, 3.9);
            \coordinate (C2) at (4.2, 3.9);
            
            \filldraw (A1) circle (0.05);
            \filldraw (B1) circle (0.05);
            \filldraw (C1) circle (0.05);

            \filldraw (4, 5) circle (0.05);
            \filldraw (4, 5.75) circle (0.05);

            \filldraw (2.6, 2.9) circle (0.05);
            \filldraw (3.2, 3.4) circle (0.05);
            \filldraw (2.1, 2.5) circle (0.05);

            \filldraw (5.4, 2.9) circle (0.05);

            \draw[decoration={brace},decorate](4.2,6.5) -- node[right] {$\gamma_A$} (4.2, 4.3);
            \draw[decoration={brace},decorate](1.5, 2.2) -- node[left=3pt, above=4pt] {$\gamma_{B_1}$} (3.6, 4);
            \draw[decoration={brace},decorate](4.4,4) -- node[right=5pt, above=2pt] {$\gamma_{B_2}$} (6.5, 2.25);

            \filldraw (A2) circle (0.05);
            \filldraw (B2) circle (0.05);
            \filldraw (C2) circle (0.05);

            \draw (A1)+(0,0.3) circle (0.5);
            \draw (B1)+(0,-0.3) circle (0.5);
            \draw (C1)+(0,-0.3) circle (0.5);

            \draw (4, 6.9) node {$A$};
            \draw (1.5, 1.6) node {$B_1$};
            \draw (6.5, 1.6) node {$B_2$};
            
        
        
            \draw [dashed](A1)--(A2);
            \draw [dashed](B1)--(B2);
            \draw [dashed](C1)--(C2);
            
            \draw (4,4) circle (0.5);
            \draw (4,3.5) node[below]{$u$};
        

            
        
        
            
\end{tikzpicture}

    {\small (a) }
    \end{minipage}
    \begin{minipage}{0.24\textwidth}
    \centering
        \begin{tikzpicture}[scale=0.8]

            \coordinate (A1) at (4, 6.5);
            \coordinate (B1) at (1.5, 3.8);
            \coordinate (C1) at (4.3, 3.5);
            \coordinate (A2) at (4, 4.2);
            \coordinate (B2) at (3.6, 3.8);
            \coordinate (C2) at (4, 3.8);
            
            \filldraw (A1) circle (0.05);
            \filldraw (B1) circle (0.05);

            \filldraw (4, 5) circle (0.05);
            \filldraw (4, 5.75) circle (0.05);

            \filldraw (2.6, 3.8) circle (0.05);
            \filldraw (3, 3.8) circle (0.05);
            \filldraw (2.1, 3.8) circle (0.05);


            \draw[decoration={brace},decorate](4.2,6.5) -- node[right] {$\gamma_A$} (4.2, 4.3);
            
            \draw[decoration={brace},decorate](B1)+(0,0.2) -- node[left=3pt, above=4pt] {$\gamma_{B_1}$}(3.6,4);
            

            \filldraw (A2) circle (0.05);
            \filldraw (B2) circle (0.05);


            \draw (4, 6.9) node {$A$};
            \draw (4,6.8) circle (0.5);
            \draw (1.1, 3.7) node {$B_1$};
            \draw (1.2,3.7) circle (0.5);
            \draw (4.2, 3.8) node {$B_2$};
            
        
        
            \draw [dashed](A1)--(A2);
            \draw [dashed](B1)--(B2);
            
            \draw (4,3.8) circle (0.7);
        

            
        
        
            
\end{tikzpicture}
\vspace{0.3cm}

{\small (b) }
    \end{minipage}
    \caption{\small 3-GHZ state distribution.}
    \label{fig:3ghz-dist}
\end{figure}

After Phase 1 of link-level bipartite entanglement distribution, the goal of Phase 2 is to establish a 3-GHZ state among the three terminal nodes. In general, this can be achieved  as shown in Fig.~\ref{fig:3ghz-dist}a. Specifically, 
a center node, $u$, has a path with each of the three terminal nodes, and on each path, link-level bipartite entanglement has been established successfully. Then for each path between a terminal node and $u$, the quantum repeaters along the path performs BSM 
to establish an end-to-end entanglement between the terminal node and $u$. After that, $u$ performs a 3-GHZ projective measurement 
so as to create a 3-GHZ state among the three terminal nodes. 

A special case of the above is that the center node $u$ overlaps with a terminal node, forming the graph in Fig.~\ref{fig:3ghz-dist}b. In this case, the overlapping terminal node will perform an entanglement fusion measurement to create a 3-GHZ state among the three terminal nodes.

Consider the setting in Fig.~\ref{fig:3ghz-dist}a. Denote the path from a terminal node $T$ to the center $u$ as $P=(T, v_1, \ldots, v_{k}, u)$, where $v_i$ is a quantum repeater. Then $T$ and $u$ share a state as in Eq. (\ref{eq:ms-final-state}).
Let $\gamma_A$ be the  parameter for the path between user $A$ and $u$.
Similarly, let $\gamma_{B_i}$  denote the parameters for the path between user $B_i$ and $u$, $i=1,2$. In the special case in Fig.~\ref{fig:3ghz-dist}b, we have $\gamma_T=1$.  

\smallskip
\noindent{\bf Estimating error rates.} To estimate the key rate in Eq. (\ref{eq:n-party-keyrate}), we need to  estimate the $Z$ basis error rate $Q_{A,B_i}$ and $X$ basis error rate $Q_{X}$. For the 3-party case, we derive them  as 
\begin{equation}
    Q_{A,B_i} = \frac{1 - \gamma_A \gamma_{B_i}}{2}\ ,
    \label{eq:Q_AB_i}
\end{equation}
\begin{equation}
    Q_X = \frac{1 - \gamma_A \prod_i \gamma_{B_i}}{2}\ .
    \label{eq:Q_X}
\end{equation}
The derivation is based on the stabilizer formalism and is found in the Appendix. The estimates of $Q_{A,B_i}$ and $Q_{X}$ will be
used in routing and leader selection in the next subsection. 

\subsection{Routing for 3-party QKD} \label{sec:3-party-routing}

Given the above, the routing for 3-party QKD is to find the optimal center so that the number of secret keys is maximized. This involves solving two problems: (i) for a given center node, find the optimal leader, $A$, so that the key rate is maximized, and (ii) find the optimal center node so that the key rate for the optimal leader is maximized. 

\subsubsection{Leader Decision}
Given the paths to a center, we can compute the key rate for that center and which terminal node should be the leader. The optimal leader for a given center node is the leader that results in the highest key rate.
To determine the leader, we  first consider each terminal node as the leader $A$, and compute $\max_i H(Q_{A,B_i})$ for our choice of leader. Note that we do not need to compute the entire key rate for each possible leader, since $Q_X$ is independent of leader choice. To find the leader that yields the highest key rate, we choose $A$ such that $\max_i H(Q_{A,B_i})$ is minimized. This results in the maximum key rate for that center as 

\begin{equation}
    r_{\max} = 1 - H(Q_X) - \min_A \max_i H(Q_{A,B_i})
    \label{eq:max-key-rate}
\end{equation}

\subsubsection{Expected key rate for a given center}
We wish to choose the center that will yield the highest key rate. Using the asymptotic key rate equation ~\eqref{eq:max-key-rate}, we can determine how many keybits are distributed per noisy 3-GHZ state. However since our link generation and entanglement swapping operations are probabilistic, it is not guaranteed that a noisy 3-GHZ state will be distributed. To account for the probabilistic nature of our network, we obtain expected key rate per round.
To accomplish this we compute the probability that our noisy 3-GHZ state has been distributed. 
Specifically, this probability 
depends solely on whether our entanglement swapping operations were successful (since this is in Phase 2, after link-level entanglement establishment in Phase 1). The number of entanglement swapping operations performed is proportional to the lengths of the paths to our center node. Consider the path from terminal $T_i$ (which can be $A$, $B_1$ or $B_2$) to the center. Let the length of this path be $\ell_i$. Establishing end-to-end entanglement along this path requires $\ell_i-1$ entanglement swapping operations. This results in terminal $T_i$ sharing entanglement with the center node with probability $q^{\ell_i-1}$. We can repeat this for the set of terminals, $\{T_i\}$, connected to the center, with an additional $q$ for the 3-GHZ measurement (or entanglement fusion) at the center node. 
Summarizing the above, the expected effective key rate for the given center is 
\begin{equation}
   r_t =  q\left[\prod_{i \in T} q^{\ell_i-1} \right] \left[1 - H(Q_X) - \min_A \max_i H(Q_{A,B_i})\right]
    \label{eq:key-rate-timestep-dynamic}
\end{equation}


\subsubsection{Center Selection}
We find the optimal center node using a method similar to the shortest-star algorithm in \cite{Bugalho_2023}. Intuitively the paths from each terminal to the best center will be the paths that yield the highest key rate. We will refer to such paths as the \emph{shortest paths}. Additionally the paths to our center node must be disjoint due to entanglement being a resource that can only be used once. This means that our paths to the best center will be the shortest possible disjoint paths. Knowing this, we can consider each node as a center and if shortest disjoint paths exist we can compute its key rate. This results in the algorithm described below.

\begin{itemize}
    \item[1.] Each terminal finds all shortest paths to each other node in the network.
    \item[2.] Each node in the network is considered as the center choice. If disjoint paths to that center do not exist, the node is ignored. Otherwise the expected key rate, $r_t$, of that center is computed (see Eq.~\eqref{eq:key-rate-timestep-dynamic}).
    \item[3.] The node that yielded the highest key rate is chosen as the center of our star.
\end{itemize}

\tydubfigsingle{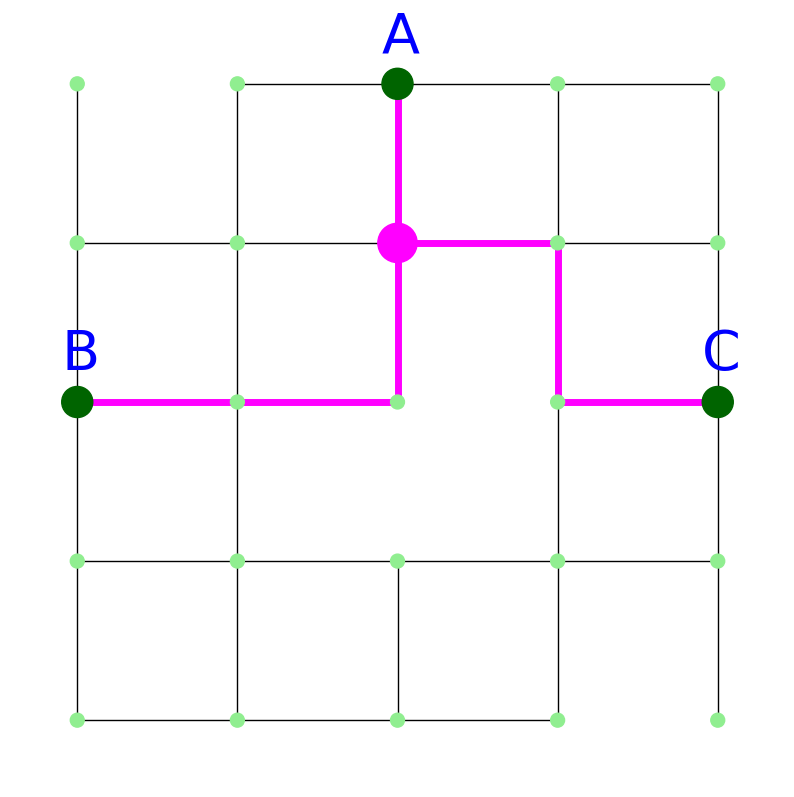}{}{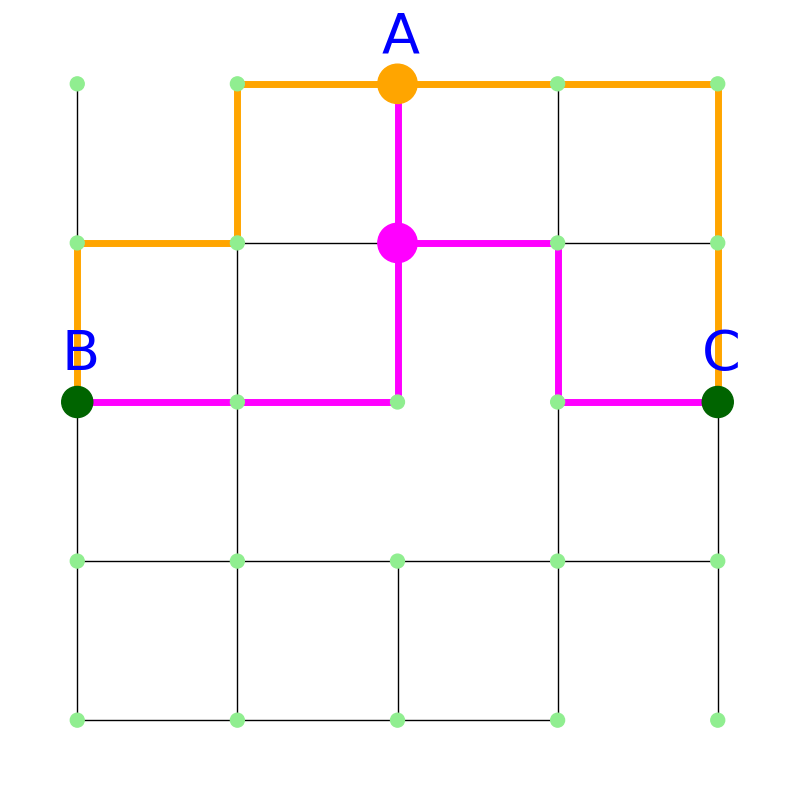}{}{(a) Running the star algorithm results in finding one star. (b) Greedily packing stars by repeatedly running the star algorithm until no more star is  found yields two stars.}{multi-star}

Running the above algorithm results in a \emph{star} with the three terminal nodes at the tip of the star. In the special case where a terminal node is the center of our \emph{star}, our graph is a path with the remaining two terminals located at each end. 
One example is shown in Fig.~\ref{fig:multi-star}a. Notably we can see that remaining entanglement exists in the residual graph after finding our star in Fig.~\ref{fig:multi-star}a. If we then remove all edges used in our star,
and re-run the star algorithm we are able to find another star as seen in Fig.~\ref{fig:multi-star}b. We  
repeat this process of greedily packing stars until no more can be found. 
Henceforth, we refer to the above algorithm as {\em multi-star} algorithm.
%



\section{n-Party QKD} \label{sec:n-party}


\begin{figure}
    \centering

    \begin{minipage}{0.49\textwidth}
    \centering
        \begin{tikzpicture}[scale=0.7]

    \coordinate (T) at (5.9,4.5);
    \coordinate (R) at (5.9,3.8);
    \coordinate (RECT) at (5.5,2.3);

    \draw[draw=black] (RECT) rectangle ++(4.2,2.6);
    \draw[draw=orange,fill=orange] (T) circle (0.25);
    \draw[draw=black,fill=lightgray] (R) circle (0.25);
    \draw[line width=0.5mm] (5.7, 3.2) -- (6.1,3.2);
    
    \draw[draw=blue,line width=0.5mm] (5.7, 2.7) -- ++(0.4,0);
    \draw[draw=red, line width=0.5mm] (5.7, 2.5) -- ++(0.4,0);

    \node[] at ($(T)+(1.3,0)$) {\small Terminal};
    \node[] at ($(R)+(1.3,0)$) {\small Repeater};
    \node[] at ($(6.4,3.2)+(1.3,0)$) {\small Network Link};
    \node[] at ($(6.4,2.6)+(1.6,0)$) {\small Distribution Path};

    \node[shape=circle,draw=orange,fill=orange,minimum size=0.7cm, line width=0.5mm] (T1) at (0,3) {\small $T_1$};
    \node[shape=circle,draw=orange,fill=orange,minimum size=0.7cm, line width=0.5mm] (T2) at (0,0) {\small$T_2$};
    \node[shape=circle,draw=orange,fill=orange,minimum size=0.7cm, line width=0.5mm] (T3) at (6,1.5) {\small $T_3$};
    \node[shape=circle,draw=orange,fill=orange,minimum size=0.7cm, line width=0.5mm] (T4) at (9,0) {\small $T_4$};

    \node[shape=circle,draw=black,fill=lightgray,minimum size=0.7cm] (I1) at (2,1.5) {\small$R_1$};
    
    \node[shape=circle,draw=blue,fill=lightgray,minimum size=0.7cm,line width=0.75mm] (I2) at (4,1.5) {\small$R_3$};

    \node[shape=circle,draw=black,fill=lightgray,minimum size=0.7cm] (I4) at (7.5,0.75) {\small$R_6$};

    \node[shape=circle,draw=black,fill=lightgray,minimum size=0.7cm] (I5) at (6,0) {\small$R_5$};
    \node[shape=circle,draw=black,fill=lightgray,minimum size=0.7cm] (I6) at (4,0) {\small$R_4$};
    \node[shape=circle,draw=black,fill=lightgray,minimum size=0.7cm] (I7) at (2,0) {\small$R_2$};


    \path [-,color=blue, line width=0.75mm] (T1) edge (I1);
    \path [-] (T2) edge (I1);
    \path [-,color=blue, line width=0.75mm] (I1) edge (I2);
    \path [-,color=blue, line width=0.75mm] (I2) edge (T3);
    \path [-] (T3) edge (I4);
    \path [-,color=blue, line width=0.75mm] (I4) edge (T4);
    \path [-,color=blue, line width=0.75mm] (I4) edge (I5);
    \path [-,color=blue, line width=0.75mm] (I5) edge (I6);
    \path [-,color=blue, line width=0.75mm] (I6) edge (I2);
    \path [-,color=blue, line width=0.75mm] (T2) edge (I7);
    \path [-,color=blue, line width=0.75mm] (I2) edge (I7);

\end{tikzpicture}

    {\small (a) Multipartite entanglement distribution via a center node (star).}
    \end{minipage}

    \begin{minipage}{0.49\textwidth}
    \centering
\begin{tikzpicture}[scale=0.7]
    \node[shape=circle,draw=orange,fill=orange,minimum size=0.7cm, line width=0.5mm] (T1) at (0,3) {\small $T_1$};
    \node[shape=circle,draw=orange,fill=orange,minimum size=0.7cm, line width=0.5mm] (T2) at (0,0) {\small $T_2$};
    \node[shape=circle,draw=orange,fill=orange,minimum size=0.7cm, line width=0.5mm] (T3) at (6,1.5) {\small $T_3$};
    \node[shape=circle,draw=orange,fill=orange,minimum size=0.7cm, line width=0.5mm] (T4) at (9,0) {\small $T_4$};

    \node[shape=circle,draw=black,fill=lightgray,minimum size=0.7cm] (I1) at (2,1.5) {\small $R_1$};
    
    \node[shape=circle,draw=black,fill=lightgray,minimum size=0.7cm,] (I2) at (4,1.5) {\small $R_3$};

    \node[shape=circle,draw=black,fill=lightgray,minimum size=0.7cm] (I4) at (7.5,0.75) {\small $R_6$};

    \node[shape=circle,draw=black,fill=lightgray,minimum size=0.7cm] (I5) at (6,0) {\small $R_5$};
    \node[shape=circle,draw=black,fill=lightgray,minimum size=0.7cm] (I6) at (4,0) {\small $R_4$};
    \node[shape=circle,draw=black,fill=lightgray,minimum size=0.7cm] (I7) at (2,0) {$R_2$};



    \path [-,color=red, line width=0.75mm] (T1) edge (I1);
    \path [-,color=red, line width=0.75mm] (T2) edge (I1);
    \path [-,color=red, line width=0.75mm] (I1) edge (I2);
    \path [-,color=red, line width=0.75mm] (I2) edge (T3);
    \path [-,color=red, line width=0.75mm] (T3) edge (I4);
    \path [-,color=red, line width=0.75mm] (I4) edge (T4);
    \path [-] (I4) edge (I5);
    \path [-] (I5) edge (I6);
    \path [-] (I6) edge (I2);
    \path [-] (T2) edge (I7);
    \path [-] (I2) edge (I7);
\end{tikzpicture}

    {\small (b) Multipartite entanglement distribution via a tree.}
    \end{minipage}
    
    \caption{\small Difference between star distribution and tree distribution visualized. (a) 4-GHZ state distributed via a central node highlighted in blue. The links that make up the paths to the center node are highlighted in blue. (b) 4-GHZ distributed via a tree. The edges of the tree are colored in red.}
    \label{fig:tree-visualization}
\end{figure}
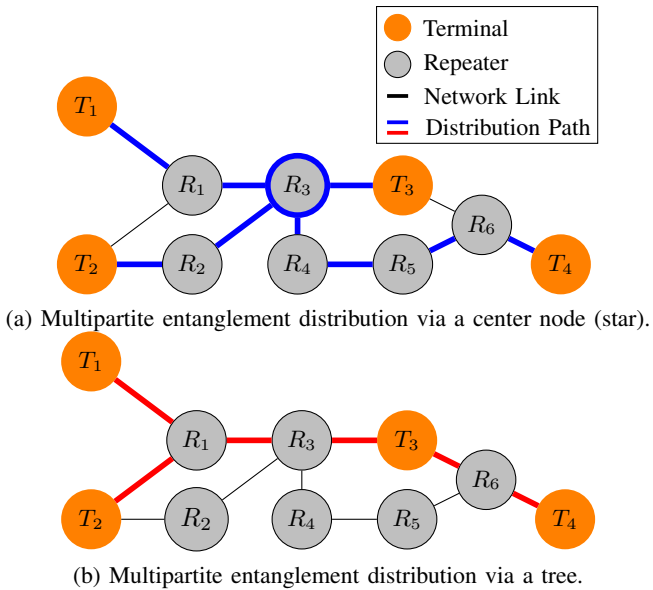

\subsection{$N$-GHZ State Distribution}
For general $N$-GHZ 
 state distribution, similar to the 3-GHZ case described in Section \ref{sec:3-party}, the goal after Phase 1 is to establish an $N$-GHZ state amongst our $N$ terminal nodes. As shown in \cite{Meignant19:graph-state}, one way to distribute an $N$-GHZ state in a network setting is through a tree. Specifically, a  tree may consist of {\em leaf} nodes (degree of 1), {\em path} nodes (degree of 2), and {\em branch} nodes (degree of $\geq$3). In Fig.~\ref{fig:tree-visualization}b, we see that $T_1, T_2,$ and $T_4$ are leaf nodes, $T_3$ is a path node, and $R_1$ is an branch node. If our tree consists of a single branch node and no terminal path nodes, we have the same case as the star described in Section \ref{sec:3-party-routing}. This can be seen in Fig.~\ref{fig:tree-visualization}a, where $R_3$ is the only branch node and will perform a 4-GHZ projective measurement. Otherwise, each path node will perform Bell state measurements (entanglement swapping) to establish bipartite entanglement between branch nodes and between leaf and branch nodes. Branch nodes will perform $k$-GHZ projective measurements to create a $k$-GHZ state between the connected nodes. Repeating this process will result in a $N$-GHZ state shared among the $N$ terminal nodes. Similar to the 3-party case, a special case occurs when an branch or path node is also a terminal node. In this case, the terminal node will perform an entanglement fusion operation resulting in a $(k+1)$-GHZ state where the terminal node retains an entangled qubit. This can be seen in Fig.~\ref{fig:tree-visualization}b, where $T_3$ is a path node, and $T_3$ will perform entanglement fusion.

\subsection{Routing for N-party QKD} \label{sec:n-party-routing}

\begin{figure*}[t]
\centerline{
    \begin{minipage}{0.24\textwidth}
      \begin{center}
        \leavevmode
        \setlength{\epsfysize}{0.85\textwidth}
        \epsffile{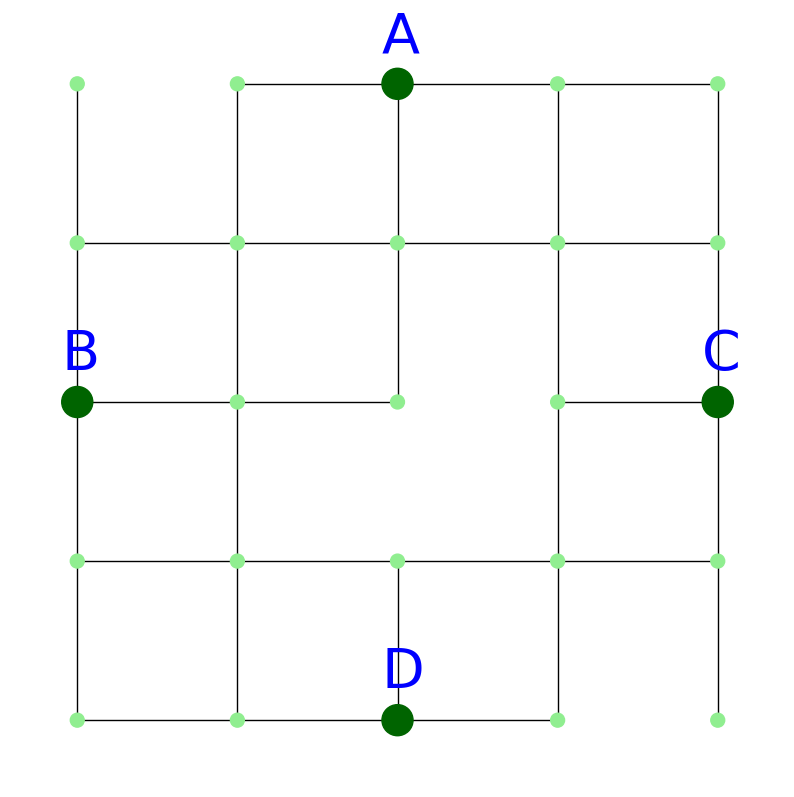}\\
       {\small (a) Initial snapshot of $G$.}
      \end{center}
    \end{minipage}
    \hspace{0.2in}
    \begin{minipage}{0.24\textwidth}
      \begin{center}
        \leavevmode
        \setlength{\epsfysize}{0.85\textwidth}
        \epsffile{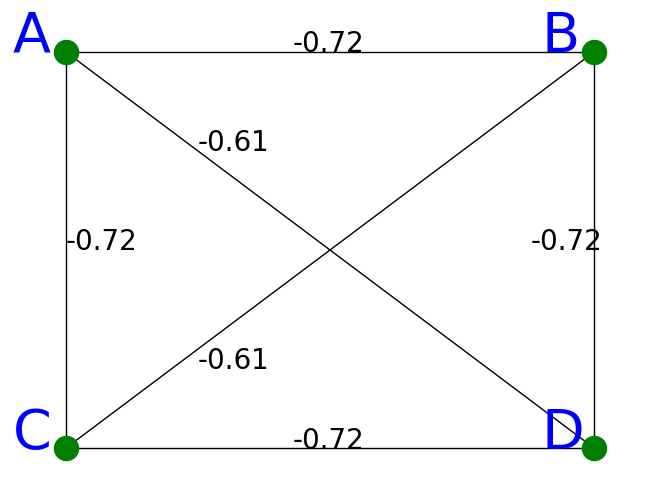}\\
       {\small (b) Subgraph $G_1$.}
      \end{center}
    \end{minipage}
    \hspace{0.2in}
    \begin{minipage}{0.24\textwidth}
      \begin{center}
        \leavevmode
        \setlength{\epsfysize}{0.85\textwidth}
         \epsffile{\Figdir 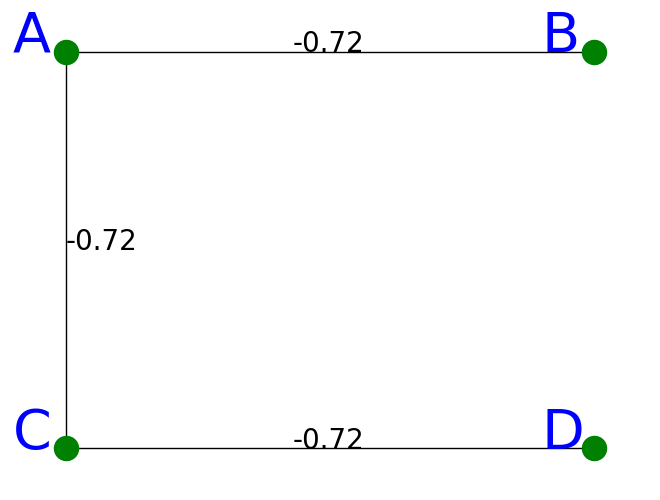}\\
       {\small (c) Min. spanning tree of $G_1$.}
      \end{center}
    \end{minipage}
    \hspace{0.2in}
    \begin{minipage}{0.24\textwidth}
      \begin{center}
        \leavevmode
        \setlength{\epsfysize}{0.85\textwidth}
        \epsffile{\Figdir 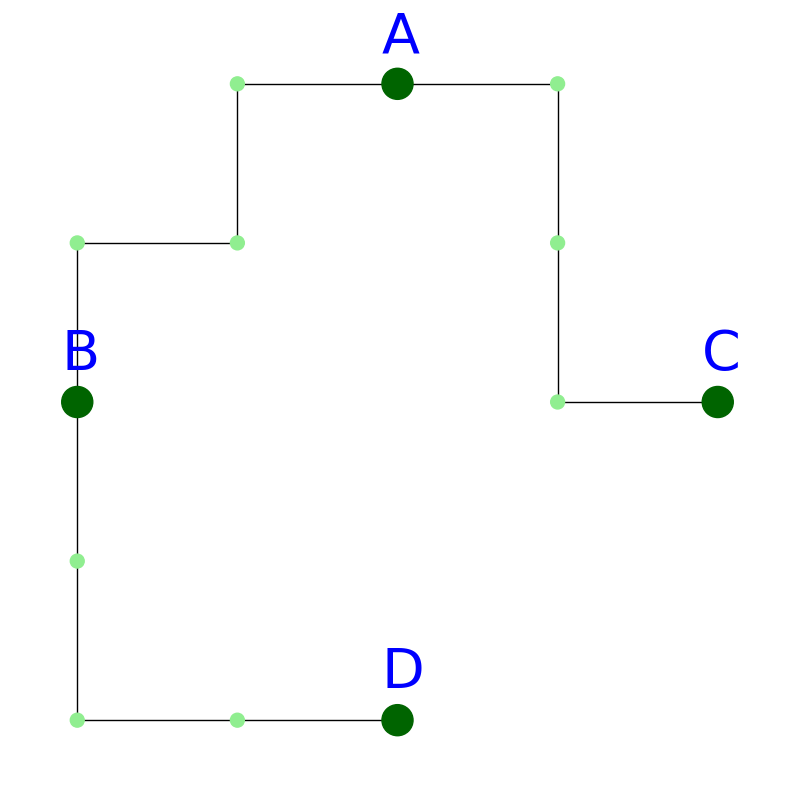}\\
       {\small (d) Outputted Steiner Tree}
      \end{center}
    \end{minipage}
} \vspace{-0.05in}\caption{\small Illustration of Steiner tree approximation algorithm used for routing for $N$-party QKD. (a) Given a snapshot of our graph $G$, first we find the shortest path between each pair of terminals. (b) Next construct the subgraph of $G$, where each edge is the shortest path between terminals. (c) Find the minimum spanning tree of $G_1$. (d) Replace each edge in $G_1$ by its corresponding shortest path.}
\label{fig:steiner-alg}
\end{figure*}


While the routing problem for 3-party QKD is finding stars with terminal nodes at the tips of the stars (see Section~\ref{sec:3-party-routing}), this may  
no longer be the best strategy for the general $N$-party case. For example, if there are (say) 5 terminals in a grid topology, it is not even possible to connect the nodes via a central node since each node has at maximum a degree of 4. Instead, we must find the best tree to distribute entanglement, where terminal nodes may be interior nodes (i.e., path or branch nodes) in our tree. 

Classically finding the minimum cost tree that connects $N$ terminals is an NP-Hard problem known as the Steiner Tree problem \cite{FastSteinerTree}. In the classical setting, the cost of the tree is determined by the sum of the weights of the paths that make up the tree. However, in our case the key rate also depends on the branch nodes, as different $k$-GHZ projective measurements/entanglement fusion operations (depending on the degree of the branch node) will result in different noisy states. Determining how branch nodes affect the fidelity, and ultimately the key rate of the final system is not trivial. Since we were unable to determine an analytical expression, we modeled these GHZ projective measurements and entanglement fusion operations using density matrices in our simulation. From these density matrices we were able to obtain $Q_{A,B_i}$ and  $Q_X$ for a given leader, $A$. Leader selection is the same as before, i.e., we iterate through all choices of $A$ and select the best one (see \S\ref{sec:3-party-routing}). As before, given a tree and the key rate for that tree we can calculate the expected key rate in a network round. 
To determine the probability that entanglement swapping is successful we count the number of non-leaf nodes in our tree, $n$. We then compute $q^{n}$, the probability that all swaps are successful. This yields an equation similar  to Eq.~\eqref{eq:key-rate-timestep-dynamic}. In the special case where the tree is a star, we can use similar expressions to the ones previously shown in Eq.~\eqref{eq:Q_AB_i} and~\eqref{eq:Q_X}.

The study in \cite{Sutcliffe_2023} presents multipath routing algorithms for distributing $N$-GHZ states, utilizing a Steiner tree approximation algorithm to find the best tree to distribute entanglement. We chose a similar approach and used the same approximation algorithm \cite{FastSteinerTree}. However, there is a key difference between our implementations of the approximation algorithm. The study in \cite{Sutcliffe_2023} assumes the fidelity of all links to be perfect, and focuses on routing for optimizing entanglement rate. As such, their approximation algorithm weights paths only by the number of links that they contain. Our design weights edges and paths proportional to their key rate, allowing us to find an approximate tree that is better suited for QKD. 

Our design based on the Steiner tree approximation algorithm\cite{FastSteinerTree} consists of the following five main steps (see illustration in Fig.~\ref{fig:steiner-alg}): 
\begin{itemize}
    \item[1.] Find the shortest path between each pair of terminals. 
    %
    Construct the subgraph $G_1$ where each vertex is a terminal, and each edge is the shortest path between these terminals, where the weight of each edge in G$_1$ corresponds to the negative key rate of the associated shortest path. 
    \item[2.] Find the minimum spanning tree of $G_1$. Since our edges are weighted by the negative key rate, the minimum spanning tree will connect terminals by paths with maximal key rates.
    \item[3.] Construct the subgraph $G_S$ where each edge of the minimum spanning tree is replaced by the shortest path.
    If $G_S$ is a Steiner tree, then we are done (i.e., we can skip steps 4 and 5). This is the case for the example illustrated in Fig.~\ref{fig:steiner-alg}. 
    \item[4.] Find the minimum spanning tree of $G_S$. 
    \item[5.] Delete edges in the minimum spanning tree until the only leaves are terminals.
\end{itemize}

\tydubfigsingle{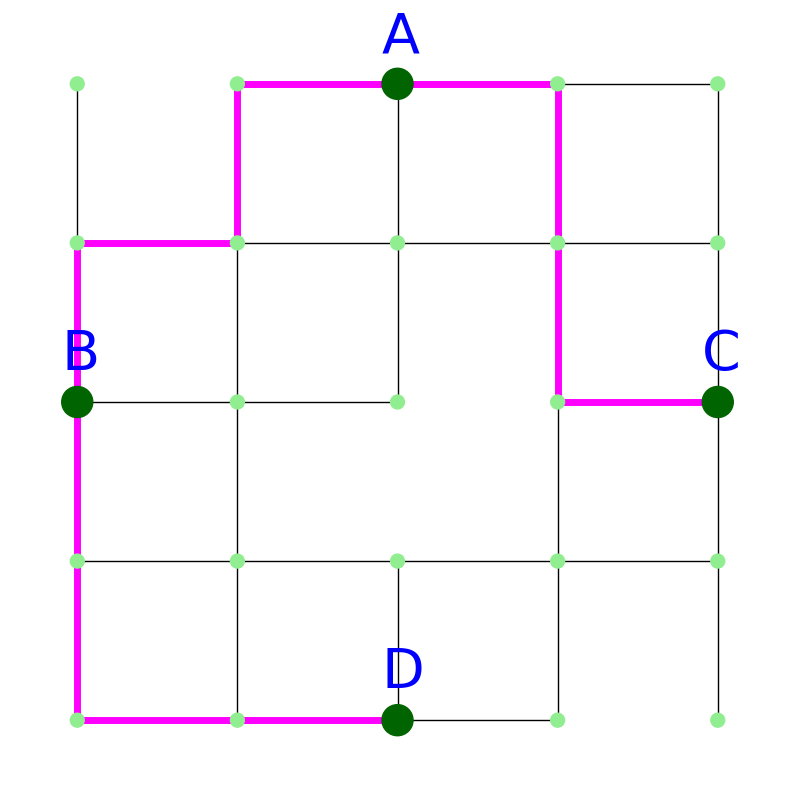}{}{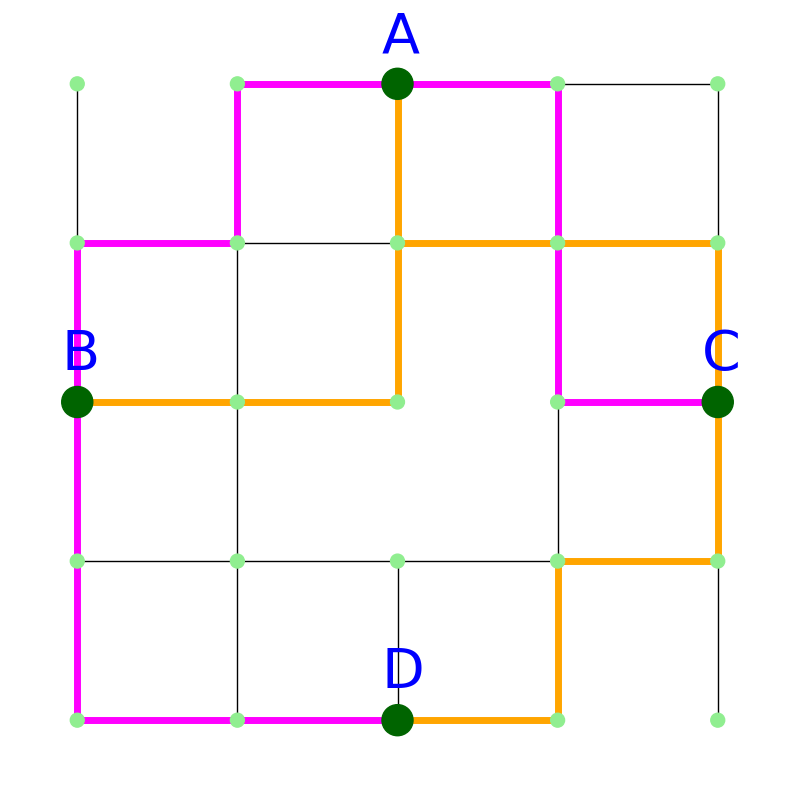}{}{Running the Steiner tree approximation algorithm results in finding one tree in (a). Greedily packing trees by repeatedly running the approximation algorithm until no trees are found yields two trees in (b).}{tree-packing}


After running our algorithm and finding a tree to distribute an $N$-GHZ state there may still be residual entanglement. One example is shown 
in Fig.~\ref{fig:tree-packing}a. We can once again repeatedly remove all edges forming our tree from the graph and repeatedly run our routing algorithm until no more trees are found. Henceforth, we refer to the above algorithm as {\em multi-tree} algorithm.  

\singlefig{figs/eval/random/keyrate_Random_Center_vs_Tree_3party.png}{Key rate results for 3-party QKD in a random graph: multi-star versus multi-tree algorithm.
}{star-vs-tree}

%
The multi-tree algorithm for the general $N$-party setting can be used for 3-party case. However, since the multi-tree algorithm uses an approximation algorithm to find trees, 
while the multi-star algorithm is specifically designed for the 3-party case, the multi-star algorithm can significantly outperform the multi-tree algorithm for the 3-party case. One example is in Fig.~\ref{fig:star-vs-tree}, which shows the key rates obtained by the two algorithms in a random graph for 3-party QKD (see more details on evaluation settings in \S\ref{sec:eval-setup}). It shows the results for two settings: $p=q=0.85$ and $p=q=0.95$, while varying $\gamma$ parameter for a link from 0.97 to 1 (see Eq. (\ref{eq:bell-depolarizing-channel})). We see a large gap between the multi-star algorithm and the multi-tree algorithm. This is due to the multi-star algorithm finding more stars, and stars with higher key rates. For $\gamma=1$, the multi-star algorithm finds 0.3 more stars/trees per round on average than the multi-tree algorithm. In our evaluation (\S\ref{sec:eval}), for ease of exposition, we mark all the results using our approach as ``multi-tree", which, for 3-party QKD, are obtained using the ``multi-star" algorithm  due to its significantly better performance than the general the multi-tree algorithm.

\section{Performance Evaluation} \label{sec:eval}

\triplefigmed{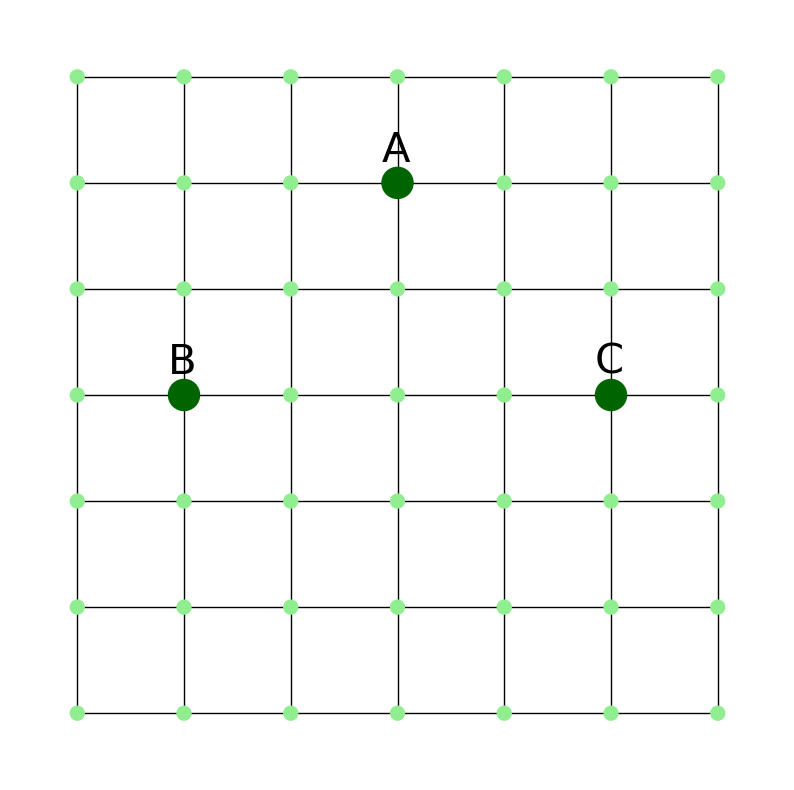}{\textphnc{b} (Bet) layout.}{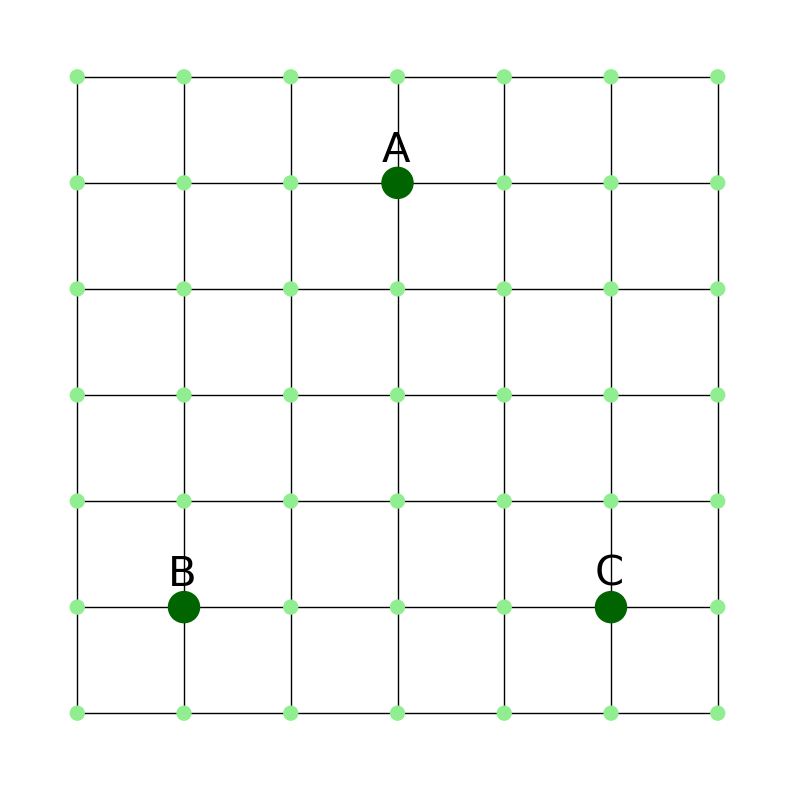}{\textphnc{d} (Dalet) layout.}{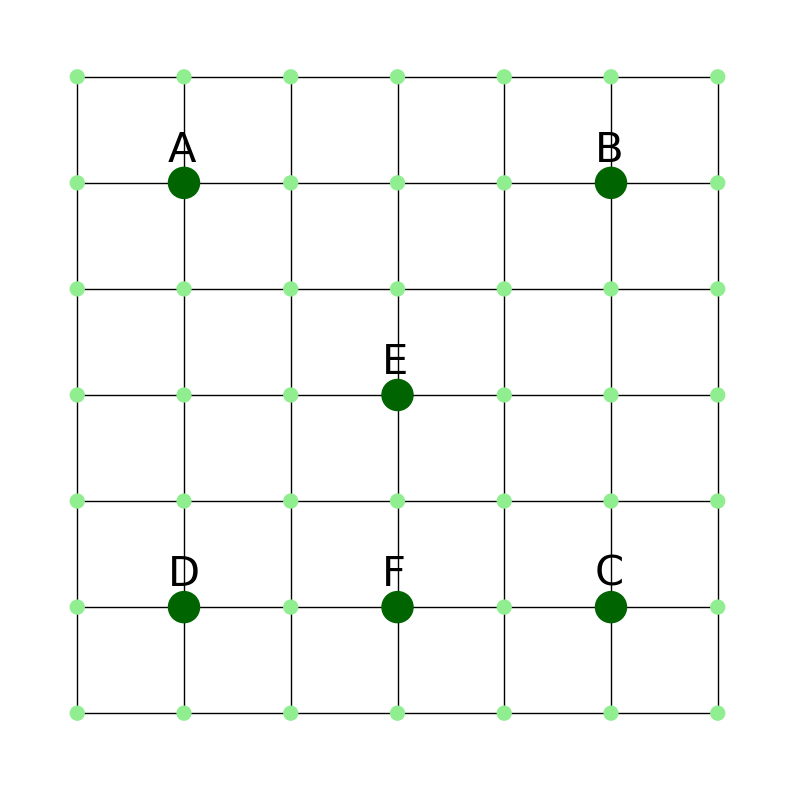}{\textphnc{g} (Giml) layout. Parties $D$, $E$, and $F$ incrementally added for $N=4,5,6$.}{Placement  of terminal nodes in a $7\times7$ grid.}{grid-layouts}

In this section, we evaluate our proposed approaches for multi-party QKD using extensive simulations. 
We first present the evaluation setup, and then the results. 

\subsection{Evaluation Setup} \label{sec:eval-setup}

We consider two different choices of routing strategies. The first choice is between \textit{fixed routing} and \textit{dynamic routing}. In \textit{fixed routing}, the routing algorithm is ran once prior to link-level entanglement being established. This results in a fixed star/tree being chosen; and if a single link in the pre-chosen path fails to generate entanglement, then entanglement distribution will fail in that round. In \textit{dynamic routing}, in each round, the routing algorithm is ran on a snapshot of the graph, where link-level entanglement success is assumed to be global knowledge. The second choice is between \textit{single-tree} vs. \textit{multi-tree} routing. In \textit{single-tree routing}, 
the routing algorithm finds a single tree for distributing entanglement, while in \textit{multi-tree routing}, trees/stars are greedily packed as described in Sections \ref{sec:3-party-routing} and \ref{sec:n-party-routing}. 
In the rest of this section, for ease of exposition, we refer to multi-star and multi-tree algorithms both as multi-tree; if it is for 3-party QKD, then the results are obtained using the multi-star algorithm.

The fixed strategies are presented as a baseline to see how dynamic strategies that operate based on global information of 
link-state success (after Phase 1) can greatly improve the key rate.
Additionally single-tree vs. multi-tree routing is intended to highlight how utilizing as many resources as possible can improve the key rate. 

Both grid and random graph topologies are examined. The grid topologies were chosen to gain insights on how varying different parameters effects the key rate. Additionally a random graph topology was chosen to see how the routing algorithms perform in a more realistic network topology.

We chose to vary several parameters and examine how they effected the key rate.
For simplicity, we consider homogeneous settings where each link/node in the network has the same parameter; our approach can be applied to heterogeneous settings. The settings we examined for these parameters are as follows: link-level entanglement success probability $p$ is set to 0.85 or 0.95,
entanglement swapping success probability $q$ is also set to 0.85 or 0.95,
and channel noise parameter $\gamma$ is varied from  0.97 to 1.0, with step size of 0.005. 
The number of parties,  $N$, is varied from 3 to 6.

\triplefignostretch{figs/eval/grid/keyrate_5x5_Bet_p85_q85.png}{\textphnc{b} (Bet) Layout}{figs/eval/grid/keyrate_5x5_Dalet_p85_q85.png}{\textphnc{d} (Dalet) Layout}{figs/eval/grid/keyrate_5x5_Giml_p85_q85.png}{\textphnc{g} (Giml) Layout}{Results for various 3-party routing schemes for three  layouts in  7$\times$7 grid,  $p=q=0.85$.}{key-rate-3party-grid}

\singlefig{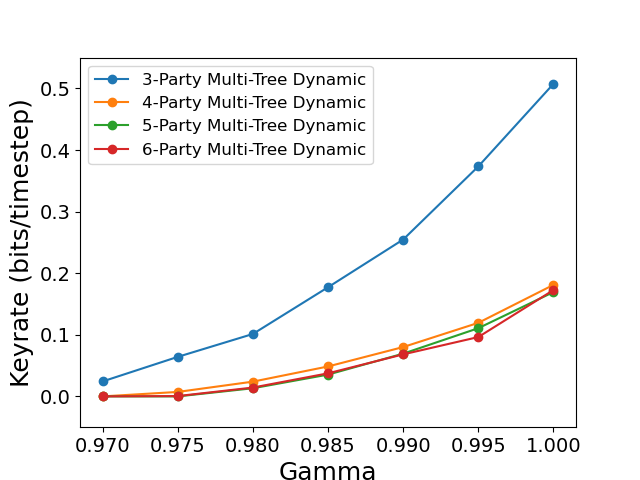}{Key rate when $N$ varies from 3 to 6 for dynamic multi-tree scheme $p=q=0.85$,  incremental \textphnc{g} (Giml) layout in 7$\times$7 grid.}{key-rate-incremental-grid}




\subsection{Results for Grid Topology} \label{sec:grid-res}

We explored two grid topologies: 7$\times$7 and 11$\times$11. For each of them, 
we examined 3 different 3-party layouts. Based on the locations of the terminal nodes, we refer to them  as \textphnc{b} (Bet), \textphnc{d} (Dalet), and \textphnc{g} (Giml) layouts due to their similarities to letters in the Phoenician alphabet. Fig.~\ref{fig:grid-layouts} shows these three layouts in the $7\times7$ grid; the layouts in the $11\times11$ grid are in a similar manner.
For both 
In 7$\times$7 and 11$\times$11 grids, we further incrementally added terminals to the \textphnc{g} (Giml) layout
to form the setup for $N=4$, 5, and 6.
%

We next present the results for the 7$\times$7 grid. In Fig.~\ref{fig:key-rate-3party-grid}, we compare how our different 3-party routing algorithms perform when $p=q=0.85$; the results for other settings show similar trend.  As expected, both the single-tree and multi-tree dynamic outperform the baselines of single-tree and multi-tree fixed respectively. We also see that the \textphnc{b} (Bet) layout yields higher key rates than the \textphnc{d} (Dalet) and \textphnc{g} (Giml) layouts. This is due to how much closer the terminals are located in the \textphnc{b} (Bet) layout when compared to the others. Interestingly, we see that the single-tree dynamic algorithm outperforms the multi-tree fixed algorithm. This is unique to the grid topology as when examining the random graph we see that multi-tree fixed algorithm outperforms the single-tree dynamic algorithm. This is partially due to the limited degree of the grid network, which makes packing trees difficult. The multi-tree fixed algorithm is able to find the maximum number of possible trees (4 trees), however the path lengths of these packed trees become very long, resulting in very long key rates for the additional trees. This significantly reduces the impact that the multi-star algorithm can have. When increasing $p$ and $q$ in a grid setting, all three layouts follow the same trends seen in the random graph (see Section \ref{sec:random-res}).




As the number of parties increases, naturally we would expect the key rate to diminish. Interestingly, for the incremental \textphnc{g} (Giml) layout we see in Fig.~\ref{fig:key-rate-incremental-grid} that 4, 5, and 6 party QKD have very similar key rates, with a large gap between the 3-party and 4-party case. The closeness of the 4, 5, and 6 party key rates due to the similarity of the trees found. The large gap between the 3 and 4 party cases can be partially explained by the non-optimality of our Steiner tree approximation algorithm (see Section \ref{sec:3-party-routing}). Additionally, this gap is exacerbated in the grid setting due to the limited connectivity between nodes. We later see that with richer connectivity, as is the case for the random graph, that the degradation for increasing $N$ is much less significant.

The results for 11$\times$11 grid show similar trends albeit with significantly lower key rates due to the increased distance between nodes. In fact for values of $\gamma$ below 0.98, all 3-party schemes in the 11$\times$11 grid result in a key rate of 0. This performance only worsens as $N$ increases.


\begin{figure*}[t]
\centerline{
    \begin{minipage}{0.24\textwidth}
      \begin{center}
        \leavevmode
        \setlength{\epsfysize}{0.85\textwidth}
        \epsffile{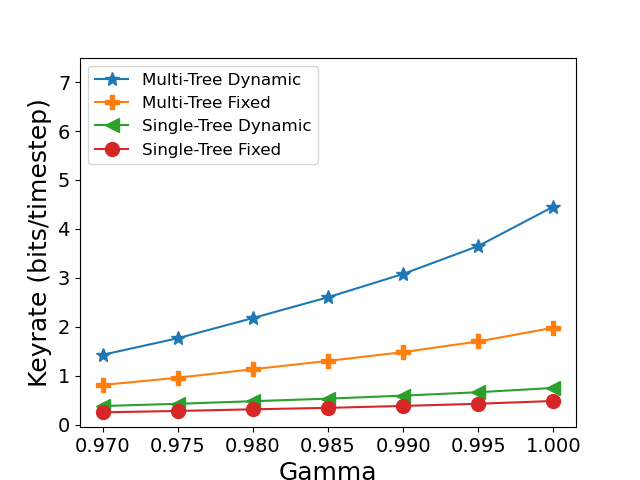}\\
       {\small (a) $p=0.85$, $q=0.85$.}
      \end{center}
    \end{minipage}
    \hspace{0.2in}
    \begin{minipage}{0.24\textwidth}
      \begin{center}
        \leavevmode
        \setlength{\epsfysize}{0.85\textwidth}
        \epsffile{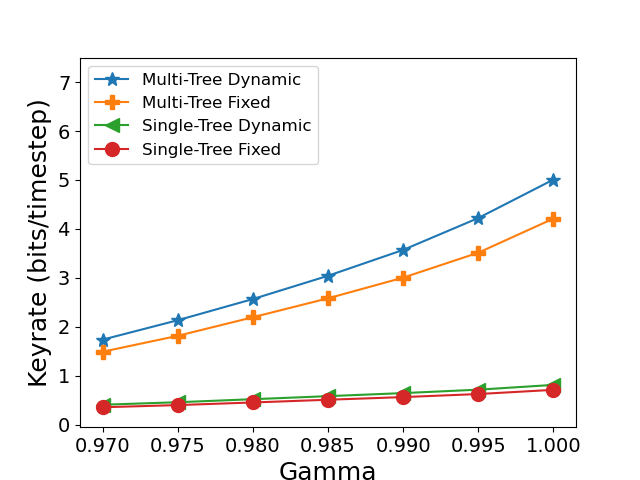}\\
       {\small (b) $p=0.95$, $q=0.85$.}
      \end{center}
    \end{minipage}
    \hspace{0.2in}
    \begin{minipage}{0.24\textwidth}
      \begin{center}
        \leavevmode
        \setlength{\epsfysize}{0.85\textwidth}
         \epsffile{\Figdir 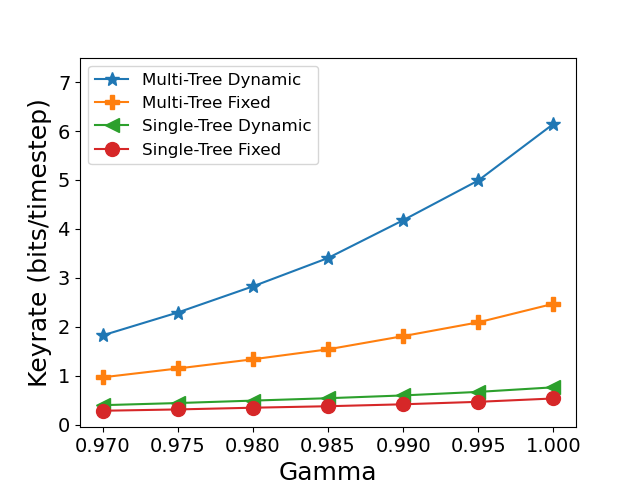}\\
       {\small (c) $p=0.85$, $q=0.95$.}
      \end{center}
    \end{minipage}
    \hspace{0.2in}
    \begin{minipage}{0.24\textwidth}
      \begin{center}
        \leavevmode
        \setlength{\epsfysize}{0.85\textwidth}
        \epsffile{\Figdir 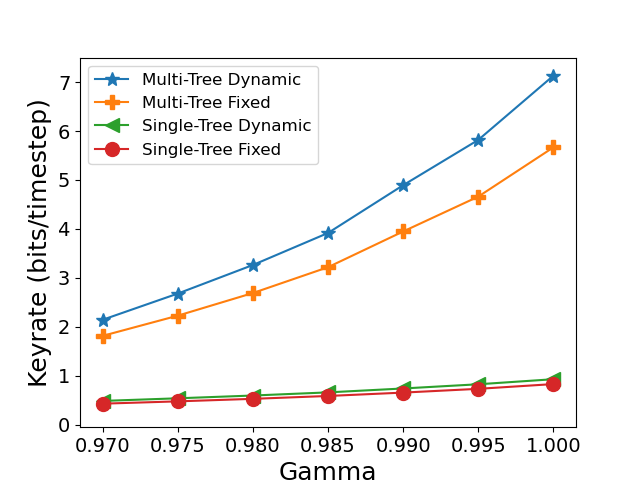}\\
       {\small (d) $p=0.95$, $q=0.95$.}
      \end{center}
    \end{minipage}
} \caption{\small Results for 3-party QKD (averaged over 5 random graphs) when varying $p$ and $q$.}
\label{fig:random-pq-results}
\end{figure*}
\begin{figure*}[t]
\centerline{
    \begin{minipage}{0.24\textwidth}
      \begin{center}
        \leavevmode
        \setlength{\epsfysize}{0.85\textwidth}
        \epsffile{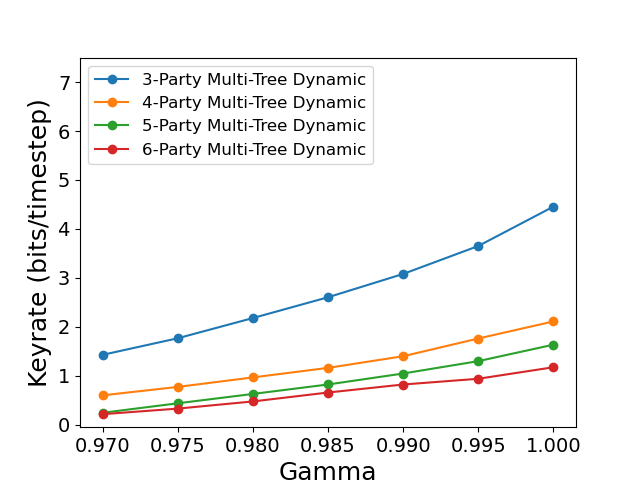}\\
       {\small (a) $p=0.85$, $q=0.85$.}
      \end{center}
    \end{minipage}
    \hspace{0.2in}
    \begin{minipage}{0.24\textwidth}
      \begin{center}
        \leavevmode
        \setlength{\epsfysize}{0.85\textwidth}
        \epsffile{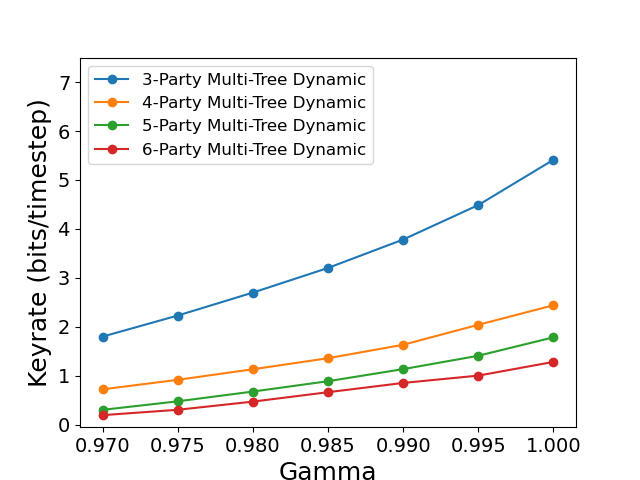}\\
       {\small (b) $p=0.95$, $q=0.85$.}
      \end{center}
    \end{minipage}
    \hspace{0.2in}
    \begin{minipage}{0.24\textwidth}
      \begin{center}
        \leavevmode
        \setlength{\epsfysize}{0.85\textwidth}
         \epsffile{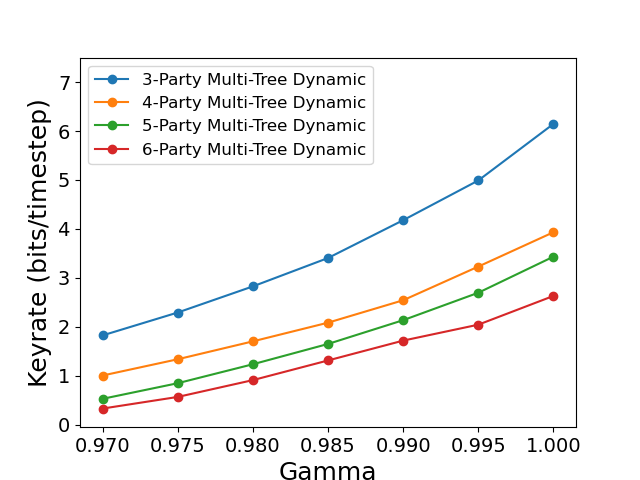}\\
       {\small (c) $p=0.85$, $q=0.95$.}
      \end{center}
    \end{minipage}
    \hspace{0.2in}
    \begin{minipage}{0.24\textwidth}
      \begin{center}
        \leavevmode
        \setlength{\epsfysize}{0.85\textwidth}
        \epsffile{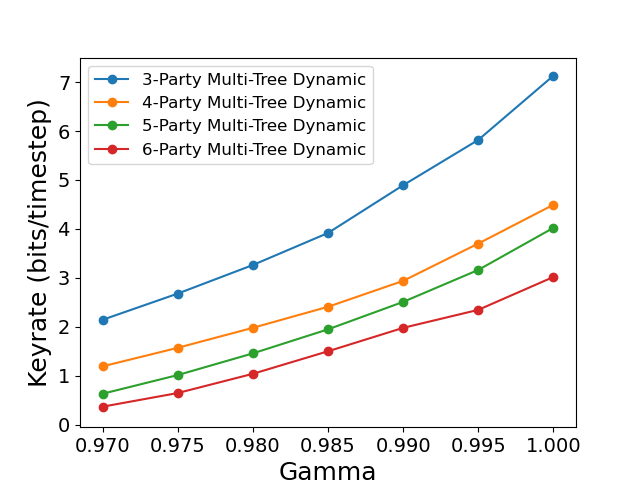}\\
       {\small (d) $p=0.95$, $q=0.95$.}
      \end{center}
    \end{minipage}
} \caption{\small Results for $N$-party QKD under dynamic multi-tree scheme (averaged over 5 random graphs) when varying $p$ and $q$, for $N=3$, 4, 5, and 6. 
}
\label{fig:key-rate-incremental-random}
\end{figure*}

\subsection{Results for Random Graph} 
\label{sec:random-res}


We examined 5 random graphs and averaged their results. Each graph is generated by placing 50 nodes randomly in a unit square. Two nodes are connected by a link if their distance is less than a radius of 0.3. These random graphs have a significantly higher average degree than our grid graphs (9.752 vs. roughly 4).
Our expectation is that this would increase the performance of the multi-tree routing, as the higher degree allows for more trees to exist.

Fig.~\ref{fig:random-pq-results} plots the average key rate for 3-party QKD in the random graphs for various values of $p$ and $q$. 
Compared to 3-party QKD for the three grid topologies in Fig.~\ref{fig:key-rate-3party-grid}, we indeed see much higher key rate for the random graphs. For instance, in Fig.~\ref{fig:random-pq-results}a, the multi-tree dynamic algorithm achieves a maximum key rate of over 4, compared to the maximum in the grid setting in Fig.~\ref{fig:key-rate-3party-grid}a of just over 0.9. 

In Fig.~\ref{fig:random-pq-results}, increasing $p$ has the most impact on fixed routing algorithms due to their reliance on specific links being established to distribute entanglement. 
Similarly, we see that increasing $q$ has the largest impact on the dynamic routing algorithms. Once again this makes sense due to $q$ and $\gamma$ being the only parameters in our key rate equation for dynamic routing. We also see that when comparing the effects of increasing $p$ and $q$ that they impact the multi-tree routing algorithms significantly more than single-tree. Intuitively this is expected since each found tree will independently be impacted by the increase of $p$ or $q$. As expected when increasing both $p$ and $q$ we see the largest increase in the key rate for both the fixed and dynamic multi-tree routing algorithms. 

We next compare the results of the different algorithms for multiparty QKD. In Fig.~\ref{fig:random-pq-results} for 3-party,   
compared to the baseline fixed single-tree  algorithm, the dynamic single-tree algorithm results in a key rate increase of $11\%$ to $54\%$ for the various values of $p$ and $q$. This increase is also seen in the dynamic multi-tree compared to the fixed multi-tree algorithm, resulting in a key rate increase of $25\%$ to $148\%$.  As we increase $N$, we see even larger percentage increase. For instance, for $N=4$, 5, and 6 parties, dynamic multi-tree algorithm leads to  $137\%$, $191\%$, $223\%$ higher key rate than fixed multi-tree algorithm when $p=q=0.85$ (figure omitted). This is because as the number of parties increases, the trees that are found 
consist of more links, which
degrades the performance of fixed routing schemes exponentially in both $p$ and $q$, whereas our dynamic schemes degrade only in $q$.

\singlefig{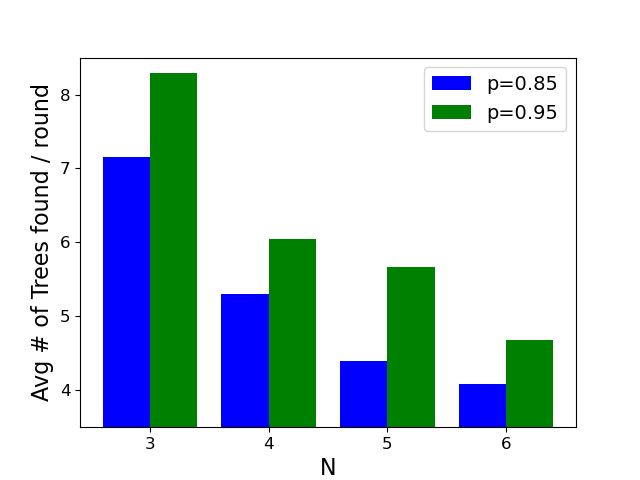}{Histogram showing the average number of trees found in each round versus $N$ when $p=0.85$ or $0.95$, and $q=0.85$.}{avg-tree-hist}

Fig.~\ref{fig:key-rate-incremental-random} plots the key rate for dynamic multi-tree algorithm when increasing $N$ in the random graphs.
We see that there is a clear gap between each plotted key rate in Fig.~\ref{fig:key-rate-incremental-random}, in contrast to the grid topology where 4, 5, and 6 parties achieved similar key rates. The incremental addition of nodes to our random graph has significantly less overlap than our incremental grid setting, which causes the trees found for increasing $N$ to be different and contain more nodes. Once again we see a large gap between the key rates as $N$ increases from 3 to 4 due to the non-optimality of our approximation algorithm. When varying $p$ and $q$ for increasing $N$ we see similar
trends exhibited in the 3-party case. We see that increasing $p$ has little effect on the dynamic routing schemes. As seen in the 3-party case, when increasing $q$, the key rate increases significantly. Notably increasing $q$ decreases the gap between the 3 and 4 party key rates.

Fig.~\ref{fig:avg-tree-hist} plots the histogram of the average number of trees found
as $N$ increases, where $q=0.85$ and $p$ is 0.85 or 0.95. We see that the average number of trees found tends to decrease when increasing $N$.
However, since we are not optimally packing trees, it is not always the case that increasing $N$ will result in less trees found, as we see is the case for $N=6$. As expected, when increasing $p$ we see that more trees are found. When more links exist in the graph, there are more possible trees that can be found. 
Finding trees is independent of $q$, which only is taken into account when attempting to distribute entanglement across trees.

\section{Related Work} \label{sec:related}

The works that are closest to ours are \cite{Meignant19:graph-state,Fischer21:graph-state,Bugalho_2023,Sutcliffe_2023}, which are on efficient multipartite entanglement distribution in quantum networks. As described in \S\ref{sec:intro}, the strategies that they design are not  for multiparty QKD. As such, their optimization goals do not match that of multiparty QKD. Our design of multi-star packing for 3-party QKD and multi-tree packing for $N$-party QKD are inspired by the above existing works. On the other hand, we explicitly consider key rate in our design to optimize  secret key generation rate for the multiple parties.  

The multipath routing strategies that we develop for multiparty QKD 
are broadly related to the extensive literature on routing in quantum networks~(e.g., \cite{caleffi2017optimal,pant2019routing,chakraborty2019distributed,Shi20:concurrent,Zhao21:Redundant,kaur2023entanglement}). 
These studies however focus on pairwise 
Bell pair distribution, instead of multipartite entanglement distribution to multiple parties. 
Specifically, these strategies use global or local information of the network, use single or multiple network paths, and are designed to improve bipartite entanglement distribution rate. 
Routing multipartite entanglement to multiple parties that is studied in this work is significantly more challenging than routing bipartite entanglement to two parties. Our designed routing strategies leverage global  information of the network; designing strategies that only use local information is left as future work.

The design and security analysis of multiparty QKD have been studied extensively (see survey in~\cite{Murta20:quantum-CKA} and the references within). These studies mainly focus on the key generation rates of a conference key agreement protocol, given an entangled GHZ state of particular fidelity.  They do not, however, consider distributing multipartite entanglement or multiparty key establishment in the context of quantum networks, which is the focus of this study. 

\mysection{Conclusion}\label{sec:concl}
In this paper, we developed 
efficient strategies for 
multiparty QKD over quantum networks. We designed strategies for both 3-party  and general $N$-party QKD  by incorporating various constraints in near-term quantum network technologies into the design to optimize secret key generation rate. Using extensive evaluation in both grid and random graphs under a wide range of settings, we demonstrate that our schemes achieve high key rate, and degrade gracefully when increasing the number of parties. 


\appendices
\section{Derivation of Error Rates for 3-party QKD} \label{app:3-QKD}
We now derive how to obtain $Q_{AB_i}$ and Q$_X$, given a star graph in a network setting.

\emph{Depolarizing Channel:} We first redefine the depolarizing channel in its general form for an $n$-qubit state $\rho$, being transmitted across channel $i$. We will need this for modelling the noise in our system. Typically the trace is omitted when defining the depolarizing channel (see Eq.~\ref{eq:bell-depolarizing-channel}) since the trace of a density matrix of a state is always 1. However it is useful here since we will be distributing the depolarizing channel over partial states.
\begin{equation}
    \mathcal{E}_i(\rho) = \gamma_i \rho + (1 - \gamma_i) tr(\rho) \frac{\mathbb{I}}{2^n}
    \label{eq:depolarizing-channel}
\end{equation}
Where $\gamma_i = \frac{4F_i - 1}{3}$, and $F_i$ is the fidelity of the state relative to the ideal state $\rho$. 

\emph{Stabilizer Formalism:} We next briefly overview the stabilizer formalism, which proves useful for modelling our noisy 3-GHZ state and taking expectation values. 
A quantum state $\ket{\psi}$ is said to be stabilized by the operator $\mathcal{U}$ if the following equation holds:
\begin{equation*}
    \mathcal{U}\ket{\psi} = \ket{\psi}\ .
\end{equation*}
The stabilizer formalism
allows us to describe quantum states by the operators that stabilize them~\cite{nielsen00}. We can then use a subgroup of the Pauli group to describe a quantum state. For the $N$-qubit state $\ket{\psi}$ stabilized by the group $\mathcal{G}$ we can describe the density matrix, $\rho$, of $\ket{\psi}$ in the following way.
\begin{equation*}
    \ketbra{\psi} = \rho = \frac{1}{2^n} \sum_{P\in \mathcal{G}}P\ .
\end{equation*}
For example, the stabilizer group for the Bell state $\ket{\Phi^+}$ is $\mathcal{G} = \{II, XX, ZZ, -YY\}$, where $\mathcal{U}\mathcal{U}$ = $\mathcal{U}\otimes\mathcal{U}$. We can then describe the density matrix as follows
\begin{equation*}
    \ketbra{\Phi^+} = \frac{1}{4} \sum_{P\in \mathcal{G}}P = \frac{1}{4}(II + XX + ZZ - YY)\ .
\end{equation*}

\emph{Error Rates for 3-party QKD:} As mentioned in \S\ref{sec:3-party}, 3-GHZ distribution for 3-party QKD is over a star topology using the aforementioned depolarizing noise model (see Eq.~\eqref{eq:depolarizing-channel}).

When modelling the noise in our star, we choose to model the noise on the outermost qubits (equivalent to modelling the noise on the center qubits). We can then perform a perfect 3-GHZ projective measurement on the center qubits. After local correction this results in a noisy 3-GHZ state where the depolarizing channel of each corresponding path is applied to each qubit of the 3-GHZ state. This is shown in the following equation: 

\begin{equation}
    \rho = \mathcal{E}_A(\mathcal{E}_{B_1}(\mathcal{E}_{B_2} (\ketbra{\text{3-GHZ}})))
    \label{eq:noisy-3ghz-initial}
\end{equation}

Using the stabilizer representation of the 3-GHZ state, we can apply the depolarizing channel and simplify (due to the trace often being zero), resulting in the final noisy 3-GHZ state shown below:
\begin{multline}
    \rho = \frac{1}{8}  (III + \gamma_A \gamma_{B_1} \gamma_{B_2} XXX  + \gamma_{B_1} \gamma_{B_2} IZZ \\ + \gamma_A \gamma_{B_1} ZZI + \gamma_A \gamma_{B_2} ZIZ - \gamma_A \gamma_{B_1} \gamma_{B_2} YYX 
     \\ -\gamma_A \gamma_{B_1} \gamma_{B_2} XYY - \gamma_A \gamma_{B_1} \gamma_{B_2} YXY)
     \label{eq:noisy-3ghz-final}
\end{multline}

For the $N$-party QKD protocol that we use in this paper (see \S\ref{sec:n-QKD}),  as defined in \cite{MP_QKD}, 
$Q_X$ is the probability that the total number of $\ket{-}$'s is odd when measuring in the $X$ basis (Note that measuring a pure $N$-GHZ state in the $X$ basis always yields an even number of $\ket{-}$'s.). Similarly $Q_{A,B_i}$ is the probability that $A$ and $B_i$ have the same measurement outcome when measuring in the $Z$ basis. That is, 
\begin{equation}
    Q_{AB_i} = \frac{1 -  \langle Z_A Z_{B_i} \rangle}{2}
    \label{eq:deriv-Q_AB_i}
\end{equation}
\begin{equation}
    Q_X = \frac{1 -  \langle X^{\otimes N} \rangle}{2}
    \label{eq:deriv-Q_X}
\end{equation}

We can then compute $Q_{AB_i}$ and $Q_X$ by taking the expectation from our noisy 3-GHZ state. Recall that the expectation of an observable, $O$, with respect to system $\rho$, is defined as $\langle O \rangle = tr(\rho O)$. Applying this we see that most of our terms can be dropped due to the trace of the (non-identity) pauli-operators being 0. For example, when taking $\langle XXX\rangle$, the only term that will have a non-zero trace is the $XXX$ term with coefficient $\gamma_A\gamma_{B_1}\gamma_{B_2}$. This results in the $Z$ and $X$ basis error rates shown in Eq.~\eqref{eq:Q_AB_i} and~\eqref{eq:Q_X}.
Additionally, these equations hold for the $N$-GHZ case, as long as the graph distributing the state is a star.

\balance
\bibliographystyle{ieeetr}
\bibliography{refs,bib/quantum2}

\end{document}